%% file: main.tex
\documentclass[longauth]{aa}

\usepackage{graphicx}
\usepackage{natbib}
\usepackage{scalerel}
\usepackage{dblfloatfix}
\usepackage{siunitx}
\usepackage{booktabs}

\usepackage[table]{xcolor}

\usepackage{listings}
\usepackage{placeins}
\usepackage[nonumberlist]{glossaries}
\makenoidxglossaries
\input{acronyms.tex}

\definecolor{codebg}{RGB}{248,248,248}
\definecolor{codeframe}{RGB}{220,220,220}
\definecolor{codecomment}{RGB}{110,110,110}
\definecolor{codekeyword}{RGB}{0,0,120}
\definecolor{codestring}{RGB}{120,20,20}
\definecolor{codenumber}{RGB}{120,120,120}

\providecommand{\Gaia}{\textit{Gaia}\xspace}

\lstdefinestyle{pythonpaper}{
    language=Python,
    backgroundcolor=\color{codebg},
    basicstyle=\ttfamily\small,
    keywordstyle=\bfseries\color{codekeyword},
    commentstyle=\itshape\color{codecomment},
    stringstyle=\color{codestring},
    numbers=left,
    numberstyle=\scriptsize\color{codenumber},
    stepnumber=1,
    numbersep=10pt,
    frame=single,
    rulecolor=\color{codeframe},
    framerule=0.6pt,
    framesep=6pt,
    xleftmargin=14pt,
    xrightmargin=4pt,
    aboveskip=0.8\baselineskip,
    belowskip=0.8\baselineskip,
    showstringspaces=false,
    showspaces=false,
    showtabs=false,
    breaklines=true,
    breakatwhitespace=true,
    keepspaces=true,
    columns=fullflexible,
    upquote=true,
    tabsize=4
}

\setacronymstyle{long-short}

\usepackage{txfonts}
\usepackage[pdfencoding=auto,psdextra]{hyperref}
\hypersetup{
    colorlinks=true,
    linkcolor=blue,
    filecolor=magenta,      
    urlcolor=blue,
    citecolor=blue
}
\makeatletter
\renewcommand*\aa@pageof{, page \thepage{} of \pageref*{LastPage}}
\makeatother

\usepackage[utf8]{inputenc}

\usepackage[switch, modulo]{lineno}
              
\renewcommand{\linenumbers}[0]{}

\usepackage{euclid}

\begin{document}
%
%

\title{Euclid Data Release 1 (DR1)}
\subtitle{An overview of the Science Ground Segment architecture
and operations}


\include{authors}

%
%
%
%


%
%
\abstract{The \Euclid mission is designed to address fundamental questions on
the nature of dark energy and dark matter by mapping the extragalactic sky with
a wide-area survey complemented by targeted deep observations. Achieving the
required scientific performance places stringent demands on the \Euclid
\gls{sgs}, responsible for transforming raw telemetry into calibrated,
validated, and scientifically exploitable data products. The \gls{sgs} is a
distributed system comprising the \Euclid Science Operations Centre, the Science
Data Centres, Instrument Operation Teams, and supporting Organisation Units,
operating within a common architectural and procedural framework. This paper
presents the \gls{sgs} as implemented for the \gls{dr1}, describing its
organisation, architecture, processing functions, and operational model. We
outline the main components of the \gls{sgs} common infrastructure. A key
element is the Euclid Archive System (EAS), which provides the central metadata
repository based on a common data model. It is composed of: a Distributed
Storage System, providing scalable storage of data files across the Science Data
Centres; a Data Processing System used by the analysis pipelines to register and
access data files and their lineage; and a Science Archive, that distributes
validated data products. Processing is coordinated through a common
orchestration framework based on the Infrastructure Abstraction Layer, which
provides workflow management and resource abstraction of the distributed compute
clusters. Software development, integration, and deployment are supported by the
Common Development Environment. We describe daily processing operations,
calibration activities, and quality control, followed by the \gls{dr1}
production campaign, including processing flows, validation strategy, and
computational and data volumes. This infrastructure enabled the production of
Euclid \gls{dr1}, involving the execution of 35 million processing jobs across
nine Science Data Centres and a total of 6.7 million core-hours, demonstrating
the robustness and scalability of the \Euclid \gls{sgs}.}
%
%
    \keywords{Physical data and processes -- Surveys -- Methods: data analysis -- Catalogues}
%
%
   \titlerunning{Euclid Science Ground Segment for DR1}
   \authorrunning{Euclid Collaboration: M.~Ruffenach et al.}
   
   \maketitle
%
%
%
%
   
\section{\label{sc:Intro}Introduction}
The \Euclid mission of the European Space Agency (ESA), conducted in partnership
with the \gls{ec}, is designed to investigate the nature of dark energy and dark
matter by mapping the geometry and growth of large-scale structure over a
substantial fraction of the extragalactic sky \citep{EuclidSkyOverview}. By
combining weak gravitational lensing and galaxy clustering probes over billions
of galaxies, \Euclid aims to deliver percent-level constraints on cosmological
parameters and to test deviations from the standard cosmological model.
Achieving these objectives requires unprecedented control of systematic effects,
uniform survey coverage, and homogeneous data processing throughout the life of
the mission. Beyond its primary cosmological objectives, \Euclid will also
enable a broad range of legacy science. These include studies of Solar System
objects and exoplanets, transient and variable phenomena, stellar populations in
the Milky Way, galaxy formation and evolution, large-scale galaxy environments,
and the identification and characterisation of rare and distant galaxies in the
early Universe.

\Euclid is conceived as a survey mission with a wide and deep observing strategy
\citep{Scaramella-EP1}, generating petabyte-scale imaging and spectroscopic data
sets thanks to the \gls{vis} instrument presented in \cite{EuclidSkyVIS} and the
\gls{nisp} described in \cite{EuclidSkyNISP}, over a nominal duration of six
years. The statistical power of the mission is directly tied to the survey area,
depth, and stability of the instrument response. Consequently, scientific return
critically depends not only on instrument performance in flight, but also on the
ability of the \gls{sgs} to process, calibrate, validate, archive, and
distribute data products with traceable quality and reproducibility.

Another defining challenge is the distributed nature of the \gls{ec}. It is
composed of numerous institutes across multiple countries, operating \glspl{sdc}
with heterogeneous computing infrastructures. The \gls{sgs} must federate these
resources into a coherent production system, ensuring consistent software
versions, data models, configuration control, and operational procedures. This
federated approach increases resilience and scalability, but requires strong
architectural principles and rigorous interface management.

The purpose of the \Euclid \gls{sgs} \citep{Pasian2012} is to transform raw
telemetry into validated, science-ready data products and to deliver them to the
\gls{ec} and the broader community according to the mission data release plan.
In operational terms, it encompasses survey planning support, instrument and
calibration data processing, routine and reprocessing pipelines, quality
control, archiving, and user support. It also provides the technical and
organisational framework that ensures traceability from pixel-level measurements
to high-level cosmological products.
\begin{figure*}
    \centering
    \includegraphics[scale=1.0]{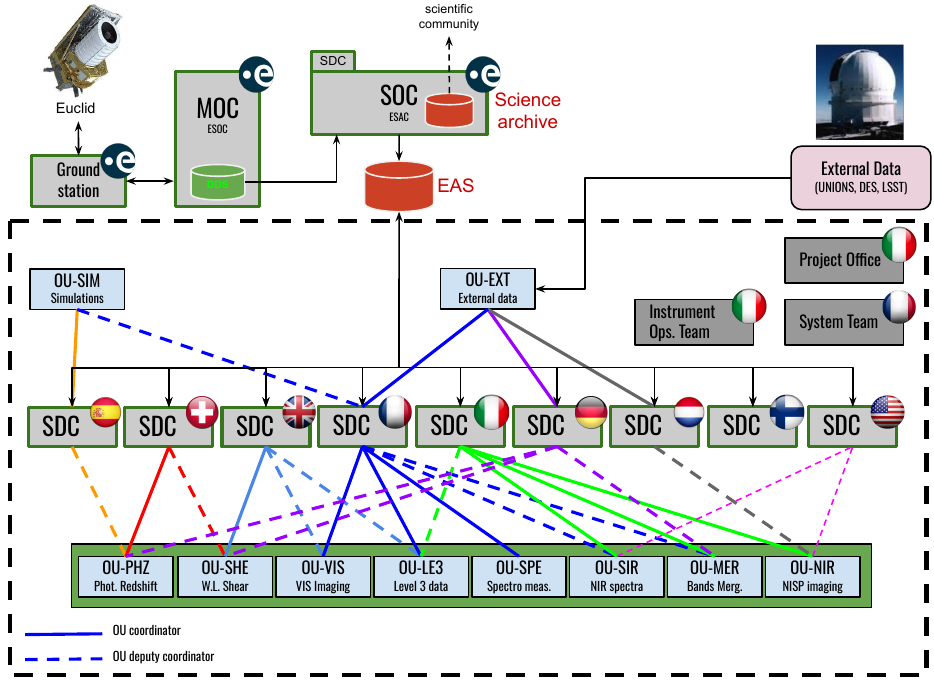}
    \caption{Organisation of the \Euclid \gls{sgs}, from \citep{Laureijs11},
    updated to reflect the current \gls{sgs} organisation. The activities inside
    the dashed box are responsibility of the \gls{ec} \gls{sgs}.}
    \label{fig:SGS_organigram}
\end{figure*}
In the following sections, this paper details the overall architecture,
processing functions, and operational model of the \Euclid \gls{sgs} as
implemented for \gls{dr1}. Given the complexity of this distributed
infrastructure and the numerous subsystems involved, a comprehensive list of the
acronyms used throughout this manuscript is provided in Appendix A for easy
reference.

\section{\label{sc:sgs}The \Euclid Science Ground Segment} 

The \gls{sgs} of the \Euclid mission \citep{Poncet2018}, illustrated in
Fig.~\ref{fig:SGS_organigram}, is composed of two independent subsystems: the
ESA \gls{soc} and the \gls{ec} \gls{sgs}. The \gls{soc} is in charge of the
survey planning, managing the downloaded data and providing raw data to the
\gls{ec} \gls{sgs}. It also serves as the single interface to the ESA \gls{moc}.
The \gls{ec} \gls{sgs} subsystem implements the processing chain that transforms
raw data into validated scientific products. While Fig.~\ref{fig:SGS_organigram}
depicts the complete ground segment organisation, this paper focuses on the
\gls{sgs} and only briefly describes the operational segment (e.g. the SOC
interface with the MOC), as needed to set the context for the \gls{sgs}
processing chain.

The \gls{ec} \gls{sgs} has the important role of producing the official \Euclid
data releases and ingesting the data products into the archive. The processing
activity is spread across nine national \glspl{sdc}, which are computing centres
responsible for implementing and running the data processing pipelines, as well
as storing the locally generated data products. The pipeline algorithms are
defined and developed by the \gls{ec} \glspl{ou}: \gls{vis}, \gls{nir},
\gls{sir}, \gls{ext}, \gls{mer}, \gls{spe}, \gls{she}, \gls{phz}, \gls{le3}, and
\gls{sim}, described in Sect. \ref{sc:pf_overview}. The \glspl{sdc} are also
responsible for instrument-related processing, and collaborate with the
\glspl{iot}, tasked with monitoring the health of the VIS and NISP instruments,
as well as their maintenance and operations.

The \gls{ec} \gls{sgs} activities are coordinated by the Project Office, which
includes the \gls{ec} \gls{sgs} Manager and a representative from the ESA
\gls{soc} \citep{Zacchei2016}. The Project Office serves as the central
management body for the ground segment, overseeing all scientific, technical and
managerial activities. It is organized around five core functional areas: system
engineering coordination, product assurance and quality assurance, configuration
control, scientific coordination, and instrument operations coordination.
Through these functions, the Project Office defines main milestones and
maintains the overall SGS schedule, coordinates end-to-end testing and
validation, manages progress reporting, facilitates cross-institutional support,
and ensures alignment with broader \gls{ec} management requirements.

The System Team defines the \gls{sgs}-wide standards and development guidelines,
covering system architectures, common tools, software interfaces, integration,
testing and validation, \gls{ec} archive structure, and data distribution within
the \gls{ec}, as well as cascading high-level \gls{ec} requirements down into
\gls{sgs}-level specifications. It is also responsible for developing the
\gls{sgs} common infrastructure described in Sect.~\ref{sc:sgs_infra}.

\begin{figure*}
    \centering
    \includegraphics[width=1\linewidth]{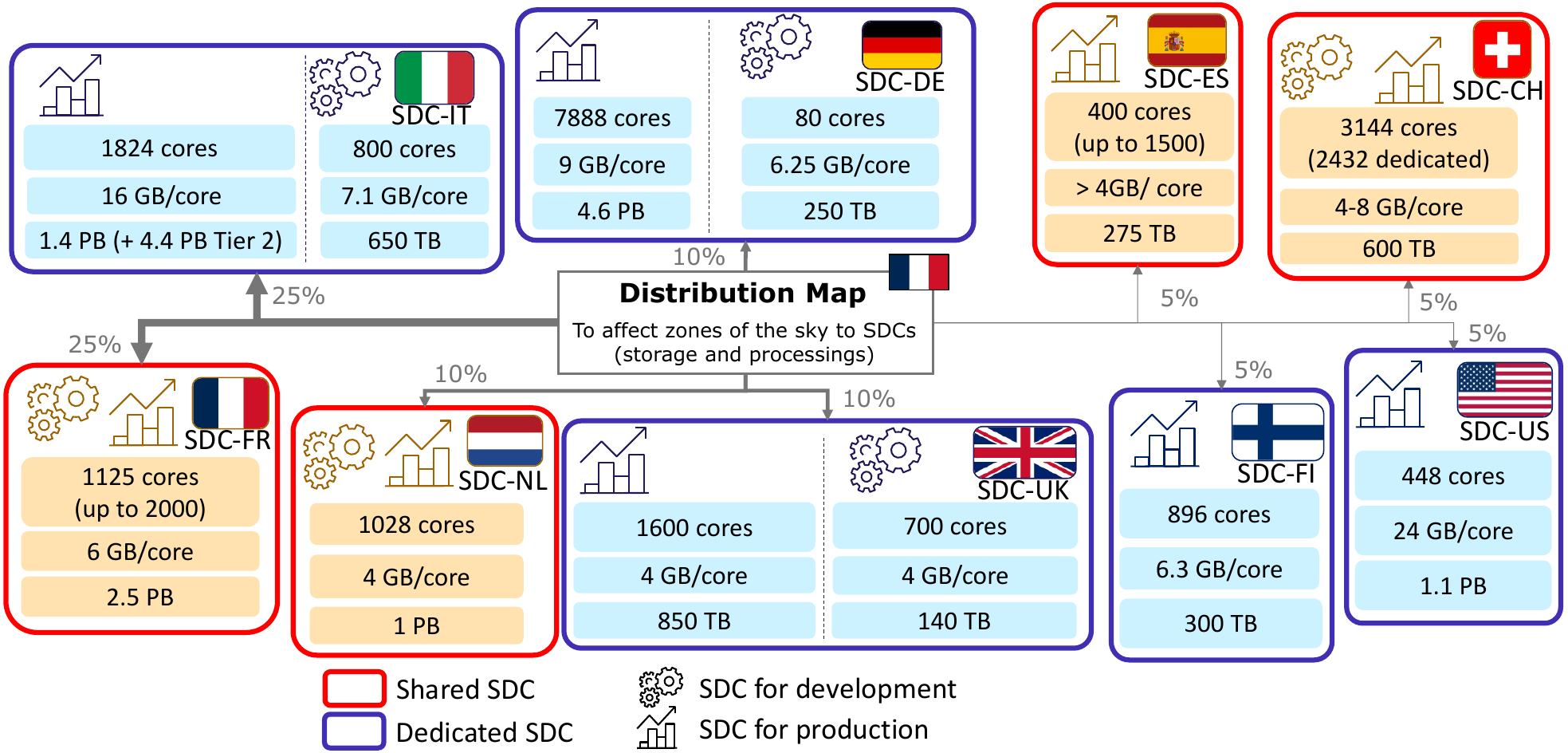}
    \caption{Resources of \gls{ec} \gls{sgs} \glspl{sdc} in terms of storage and processing resources.}
    \label{SDC_ressources}
\end{figure*}

\section{\label{sc:soc}The Science Operations Centre} 

The \gls{soc} forms the ESA Ground Segment interface between the \gls{moc} and
the \gls{ec} \gls{sgs}; principally, the \gls{soc} executes and monitors the sky
survey defined by the \gls{ec}, processes the resulting telemetry and monitors
the scientific health of the payload. 

The \gls{soc} operates survey-planning tools to enable routine
execution of the sky survey and to incorporate additional input to the mission
planning such as instrument commanding requests or additional observing
requests. It processes the payload and spacecraft telemetry daily
received from the MOC, producing the so-called \gls{le1} data products which are the
input to the \gls{ec} \gls{sgs} data processing pipelines. The \gls{le1} production
activity includes telemetry checking and assessment of both the scientific and
housekeeping telemetry and the production of daily reports.

The \gls{soc} is also responsible for designing, implementing and operating the
Euclid Science Archive, which contains the final validated \Euclid data releases, and
the Euclid Data Space, which integrates the Science Archive with visualisation and
computational services to provide seamless scientific access to the validated
data products (Sect.~\ref{sbc:eds}).

\section{\label{sc:sdc}Role of the Science Data Centres} 
\subsection{\label{sbc:sdc_dev}Software development and integration}

The \glspl{sdc} are responsible for developing \glspl{pf}, described in Sect.
\ref{sc:pf_overview}, used to generate science data products and process an
agreed subset of \Euclid survey data for each Data Release. These activities
require significant coordination both within the \glspl{sdc} and between them, to
ensure \glspl{pf} can share data products and execute tasks on the different \gls{sdc}
compute clusters. Although each \gls{sdc} is only responsible for a subset of
\gls{pf} development, as illustrated in Fig. \ref{fig:SGS_organigram}, they must
be capable of running all \glspl{pf} from start to finish on their
infrastructure.

There are nine \glspl{sdc} to process \Euclid data, namely SDC-\gls{fr},
SDC-\gls{it}, SDC-\gls{nl}, SDC-\gls{de}, SDC-\gls{uk}, SDC-\gls{ch},
SDC-\gls{fi}, SDC-\gls{es}, and SDC-\gls{us}.

SDC-FR is responsible for the development of three \glspl{pf}: VIS, SPE, and
\gls{le3}. It is co-lead for SIR, MER, and SIM. The computing infrastructure is
split into two units: SDC-FR-PROD for the production; and SDC-FR-DEV for the
development. Both rely on the existing National Institute for Nuclear and
Particle Physics Computing Centre (CC IN2P3) in Villeurbanne, France. In
addition, SDC-FR develops and hosts the \gls{codeen} platform described in Sect.
\ref{sbc:codeen}, and contributes 25\% of the processing and storage for \Euclid
observations.

SDC-IT is responsible for the development of three \glspl{pf}: SIR, MER, and
NIR. It is co-lead for \gls{le3}. It is managed by the INAF-Astronomical
Observatory of Trieste, in collaboration with the Italian Space Agency and the
ALTEC S.p.A industry. The computing infrastructure is split into two units:
SDC-IT-DEV, in Trieste, for the development; and SDC-IT-PROD for the production,
located at ALTEC S.p.A premises in Turin \citep{Bramante2025}. SDC-IT
contributes 25\% of the processing and storage for \Euclid observations.

SDC-UK is responsible for the development of the cosmic shear analysis \gls{pf}
identified as SHE, and is co-lead for VIS and \gls{le3}. The SHE pipeline is the
most computationally expensive \gls{pf} in the \gls{sgs}, accounting for
approximately 50\% of the total budget. SDC-UK is responsible for 10\% of the
processing and storage of \Euclid observations, and is operated by the Wide
Field Astronomy Unit. The processing and archiving resources are housed at the
Royal Observatory of Edinburgh. Additionally, SDC-UK is in charge of development
of the \gls{sgs} \gls{coors} described in Sect. \ref{ssbc:coors}.

SDC-DE is responsible for the development of the EXT \gls{pf} together with
SDC-NL, and is co-lead for MER. They each contribute 10\% of the processing
and storage of \Euclid observations. SDC-DE is hosted at the Max Planck
Institute for Extraterrestrial Physics in Garching, in collaboration with
the Max Planck Computing and Data Facility, and also develops and
hosts the Monitoring \& Control package described in Sect.~\ref{sbc:monit}.

SDC-NL develops and supports the \gls{dps} and \gls{dss} components of the
\gls{eas}, described in Sect. \ref{ssbc:dps} and \ref{ssbc:dss}, and implements
the \Euclid \gls{cdm}. It is hosted at the University of Groningen, with support
from Leiden Observatory. The \gls{esac} is responsible for the Science Archive
component of the \gls{eas} (Sect.~\ref{ssbc:sas}).

The remaining 20\% of computing and storage resources needed by the SGS are
equally provided by SDC-CH, SDC-ES, SDC-FI, and SDC-US \citep{teplitz2020}.

SDC-CH is responsible for the development of the PHZ \gls{pf}, as well as
developing and supporting the \gls{ial} described in Sect.~\ref{ssbc:ial}. It is
hosted at the Department of Astronomy, University of Geneva, in Versoix, and
the \gls{ial} development team is based at the FHNW University in Zurich.

SDC-ES, hosted at the Port d'Informaci\'o Cient\'ifica (PIC) in Barcelona, is
the lead of OU-SIM. SDC-FI is provided by the Cosmology Data Centre Finland
consortium, led by the University of Helsinki, and contributes to the validation
of simulations. SDC-US is operated by the Euclid NASA Science Centre at IPAC
(ENSCI), California, and contributes to the development of the NIR and SIR
\glspl{pf}.

\begin{figure*}[t]
    \centering
    \includegraphics[width=1\linewidth]{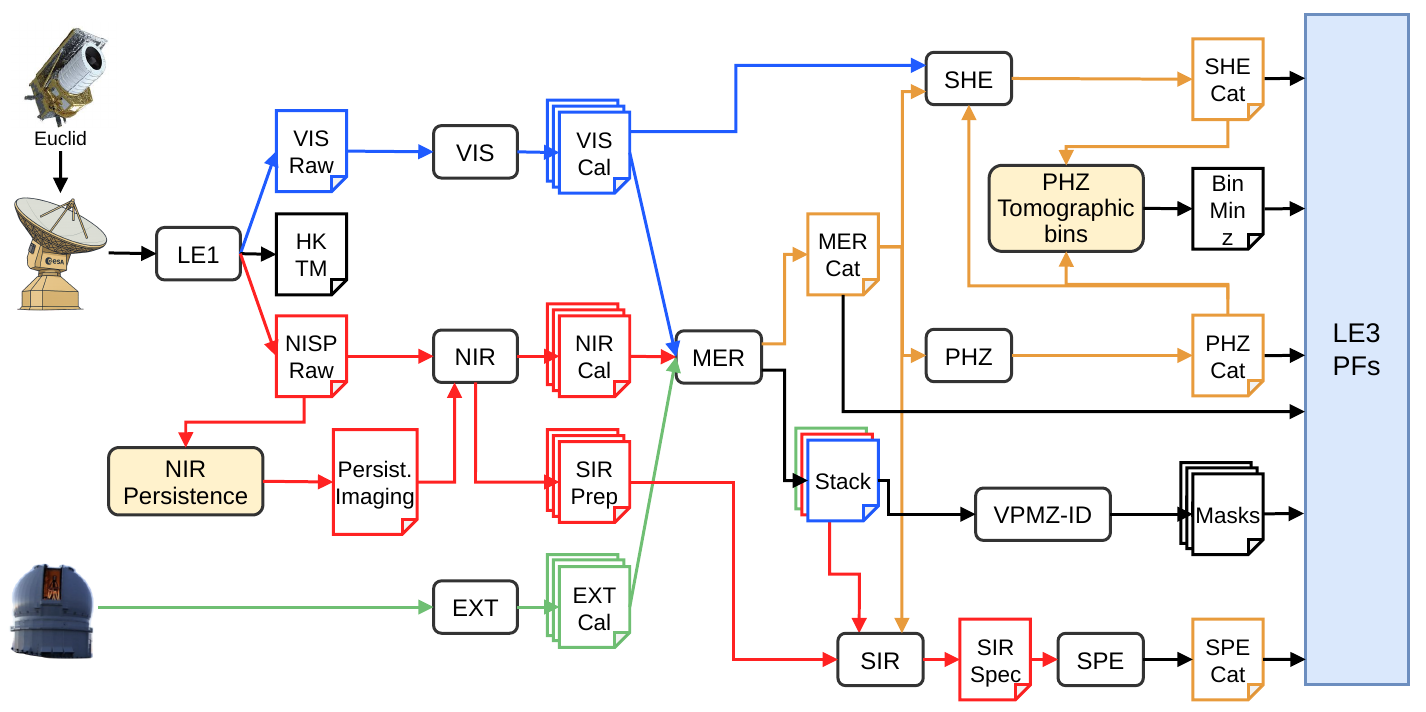}
    \caption{Illustration of the interaction of the \glspl{pf} inside the
    \gls{sgs}, as well as some key products. \glspl{pf} are depicted as rounded
    rectangles, document icons represent data products. Each \gls{pf} writes its
    outputs to the \gls{eas}-\gls{dps} before they are consumed asynchronously
    by downstream processing pipelines — no direct communication occurs between
    them.}
    \label{SGS_Pipelines}
\end{figure*}

The \gls{sdc} software development activities require taking the scientific
algorithms developed by the \glspl{ou} and implementing them through the common
development environment. This enables them to run reliably and consistently
across the heterogenous \gls{sdc} infrastructure. Developer teams mostly follow
standard industry practices for continuous integration and continuous deployment,
and the \gls{sgs} provides a common development environment that all developers
are required to utilise. The greatest challenges include coordinating efforts
between \gls{sdc} development teams to ensure \gls{pf} interoperability, and
converting scientific code bases into performant \glspl{pf} that can operate on
all the different resource platforms for both computing and storage. 

The \gls{sdc} data processing responsibilities involve deploying and operating
the \Euclid software infrastructure components (Sect.~\ref{sc:sgs_infra}) by
connecting them to local computational resources. These software infrastructure
components are used to manage the distribution of processing tasks and data
products across the \gls{sgs}. \gls{sdc} administrators operate their local
instances of these components, ensuring that the data processing executes
smoothly on their clusters and that the data products are being stored in the
local archive node. System downtimes, software component updates, and local
resource changes are communicated and coordinated with the \gls{sgs} operations
team.

\subsection{\label{sbc:sdc_prod}Production infrastructure} 
The data production infrastructure is provided by national computing facilities,
university computing centres, and/or industry involvement.
The \glspl{sdc} vary in how they deliver computing resources to perform the data
processing. Some \glspl{sdc} operate on large shared clusters, while others have
dedicated clusters specifically designed to run \Euclid workloads. This results
in a heterogeneous infrastructure with a variety of \gls{cpu} models, shared
filesystems, storage technologies, and resource footprints, as illustrated in
Fig. \ref{SDC_ressources}.

The heterogeneous infrastructure precludes use of certain technologies due to
the fact that they cannot be guaranteed at each \gls{sdc}. These include
multi-node \gls{mpi} and \glspl{gpu}, which might otherwise offer better
performance for certain \gls{pf} tasks. \gls{i/o} optimisation is also a
challenge due to different performance profiles for different shared file
systems. \gls{i/o} patterns that work well at one \gls{sdc} may perform poorly
at another, solely due to differences in the underlying storage technology. This
necessitates profiling \gls{pf} codes across multiple \glspl{sdc} before
deployment to ensure the code performs efficiently at all sites.   

The System Team estimates the overall technical budget by testing the
performance of the pipeline code and extrapolating these results to the expected
data sets for a scheduled data release. Performance metrics are measured for the
different pipelines at each \gls{sdc} to provide as accurate an estimate as
possible. Margins are added for job failures, inefficient resource usage, and
known infrastructure downtimes. \glspl{sdc} are then responsible for providing
sufficient computing resources to meet their processing/archiving commitments
according to the data release milestones. Each \gls{sdc} has its own national
process for funding and procuring computing resources, and the System Team
provides guidance to ensure overall requirements and compatibilities are
achieved across the \gls{sgs}. 

\gls{sdc} clusters need to provide access to a shared filesystem for data
workflow management and long term archive storage for processed data products.
The shared filesystems need to be optimised for high \gls{i/o}. \glspl{pf} with
particularly challenging \gls{i/o} profiles are encouraged to utilise local /tmp
disk at \glspl{sdc} which have this resource on their worker nodes. Archive
storage prioritises high volume and resilient platforms, and is not \gls{i/o}
bound. Some \glspl{sdc} are utilising tape backup for archive data.  

\section{\label{sc:pf_overview}Processing Functions overview}
In the \Euclid \gls{sgs}, data processing is organised into a set of
self-contained \glspl{pf} that interact through the central archive, as
illustrated in Fig.~\ref{SGS_Pipelines}. The processing chain begins with the
\gls{le1} \gls{pf}, which receives \Euclid satellite telemetry, acquired by the
\gls{moc} and transferred to the \gls{soc}, including VIS and NISP detector
data, \gls{hktm}, and \gls{moc} auxiliary products. \gls{le1} transforms these
inputs into VIS and NISP raw-frame products, along with \gls{hktm} products.
These products are aggregated \gls{fits} files containing decompressed detector
data, on-board timestamps converted to estimated Coordinated Universal Time, and
associated pointing information, all of which serve the basis for the subsequent
processing. From these raw data, the VIS and NIR \glspl{pf} independently
perform instrument-specific calibrations.

VIS (\citealt{Q1-TP002}; \citealt{DR1-TP002}) processes the visible imaging data
to deliver fully calibrated frames in the \IE band along with associated source
catalogues. The pipeline applies a sequence of detector-level corrections (bias
subtraction, crosstalk and non-linearity removal, brighter-fatter-effect
modelling, PRNU flattening) before cosmic-ray rejection and source detection.
Astrometric calibration uses \Gaia DR3 as reference frame, while photometric
calibration determines per-exposure zero points from quality-selected \Gaia~DR3
stars in the self-calibration field. NIR (\citealt{Q1-TP003};
\citealt{DR1-TP003}) handles the near-infrared imaging data from the 16 NISP
HAWAII-2RG detectors, producing calibrated frames in the \YE, \JE, and \HE bands
with associated catalogues. Detector-specific corrections include bad-pixel
masking, dark-current subtraction, non-linearity correction, and flat-fielding;
a key processing step is the identification and masking of persistent signal
residuals arising from charge trapping, requiring up to 5~h of exposure history.
Astrometric calibration uses \Gaia DR3 as reference catalogue, while photometric
calibration relies on a selected white dwarf observed in dedicated
self-calibration field visits. Both pipelines produce stacked frames per
\gls{ros} using \texttt{SWarp}, together with catalogues and PSF models computed
via \texttt{PSFex}.

In parallel, the EXT \gls{pf} \citep{DR1-TP006} processes complementary
ground-based observations required for photometric redshift estimation and for
constructing galaxy and stellar spectral energy distributions. These
ground-based observations come from several surveys: CFIS \citep{Ibata2017}, HSC
\citep{Miyazaki2018}, and Pan-STARRS \citep{Chambers2016} for the northern part,
which are combined through the UNIONS survey \citep{Gwyn2025}; and DES
\citep{DES2016} for the south. Single-epoch frames are calibrated, resampled
into a common MER-tile grid, and combined into stacked images using a weighted
mean. The stacking of PSF models from individual frames is performed at source
positions detected on each stacked image, using either a source-by-source or an
ensemble approach depending on field density.

The outputs of VIS, NIR, and EXT are subsequently combined by MER
(\citealt{Q1-TP004}; \citealt{DR1-TP009}), which generates stacked images
(mosaics) and consolidated photometric source catalogues. Source detection is
performed with \texttt{SExtractor++} independently on the VIS and NIR
(three-band stacked) co-added images. Photometry is measured using complementary
methods: Kron aperture, PSF-matched aperture, template-fitting, and 2D Sérsic
profile fitting. Morphological measurements include the CAS criteria
(concentration, asymmetry, smoothness) and \verb|FLUX_RADIUS|.

These merged data products are then used by PHZ (\citealt{Q1-TP005};
\citealt{DR1-TP011}) to estimate photometric redshifts and derive source
properties, and by SHE \citep{EP-Congedo} to measure galaxy shapes from the
photometric images and quantify the distortions induced by the intervening mass
distribution, dominated by dark matter, through weak gravitational lensing. The
PHZ pipeline determines photometric redshifts for all MER sources using two
complementary methods: \texttt{Phosphoros} (Bayesian SED fitting) and
\texttt{NNPZ} (nearest-neighbour algorithm using the COSMOS spectroscopic
catalogue), each providing a full photo-\texttt{z} probability density function.
Object classification into star, galaxy, and QSO classes is performed with
probabilistic Random Forests trained on a spectroscopic reference samples. In
addition, PHZ defines a set of tomographic redshift bins and for each bin it
computes the mean spectroscopic redshift. SHE \gls{pf} aims to measure the shape
of approximately 1.5 billion galaxies to identify the lensing distortion. It
measures weak gravitational lensing from galaxy shapes using a physical PSF
model of the Euclid optical channel that accounts for telescope optics,
detectors, thermal conditions, and chromatic image propagation through the
optical system. Two complementary shear estimation methods are applied: LensMC
(forward modelling of lensed galaxies described in  \citealt{EP-Congedo} and
\citealt{Congedo26}) and MetaCalibration (moment-based self-calibration
described in \citealt{Kannawadi2021}).

Spectroscopic processing follows a dedicated branch. The SIR \gls{pf}
(\citealt{Q1-TP006}; \citealt{DR1-TP023}) processes NISP slitless spectroscopy
exposures, using the MER final catalogue as the reference source catalogue to
define spectra extraction targets. SIR extracts and decontaminates source
spectra, calibrates them in wavelength and flux, and combines them into
1-dimensional spectra. These calibrated spectra are then analysed by SPE
(\citealt{Q1-TP007}; \citealt{DR1-TP025}) to measure spectroscopic redshifts. It
performs three main steps: redshift determination, using model fitting on
discrete redshift grids tailored to galaxy, star, and QSO classes; spectroscopic
classification comparing the $\chi^2$ evidence across these three categories;
emission line measurements are then performed using two independent methods,
Direct Integration on continuum-subtracted spectra and Gaussian Fitting.

At a higher level, \gls{le3} produces the mission's advanced scientific data
products. It is structured into several scientific work packages covering weak
lensing, galaxy clustering, galaxy clusters, and the construction of internally
validated data sets with well-characterised redshift selection functions for
both photometric and spectroscopic samples, as well as optimised catalogue
selections that maximise sample purity and completeness. Although \gls{le3} data
products are not included in the \gls{dr1} Foundation release, an extensive
overview of the \gls{le3} analysis pipelines, which are currently under
preparation, is provided in Appendix~\ref{sec:le3_pf}. The inclusion of this
work in an appendix reflects the fact that \gls{le3} has carried out essential
work that has also contributed to the validation of the DR1 Foundation, with
more detailed papers expected to be published with the DR1 cosmology release in
2027.

The VMPZ-ID \gls{pf} (Visibility Mask Photo-\texttt{z} Internal Data) described
in \cite{DR1-TP028} acts as the bridge between the \gls{le2} processing and the
\gls{le3} analysis pipelines. The purpose of VMPZ-ID is to assemble the survey
characterisation products that downstream \gls{le3} \glspl{pf} require to
perform statistically unbiased measurements. With increasing complexity, these
products consist of: (i) the accurate survey footprint, reconstructed from MER
tiling expressed as a \texttt{HEALPix} Multi-Order Coverage map
\citep{Gorski2005}; (ii) bit masks collected from VIS, NIR, and EXT, combined
into a unified coverage mask encoding the footprint and all non-masked regions;
(iii) instrumental and environmental information (e.g., noise maps, PSF
characterisation, Galactic extinction, and zodiacal light) needed to compute
limiting magnitudes and visibility masks; (iv) depth maps estimated from rms
noise; and (v) photometric visibility masks and random catalogues tailored to a
given source selection.

Finally, the \gls{sgs} includes a dedicated simulation pipeline, SIM
\citep{EP-Serrano}, designed to reproduce realistic observations for validation
and qualification purposes. SIM incorporates dedicated simulators for the VIS
instrument (\texttt{ELVIS}), for NISP near-infrared photometry (\texttt{Imagem},
covering the \YE, \JE, and \HE bands), for NISP slitless spectroscopy
(\texttt{TIPS}), and for external ground-based data (\texttt{SIM-EXT}), enabling
end-to-end testing of the full processing chain under controlled and realistic
observing scenarios.

\section{\label{sc:sgs_infra}SGS common infrastructure}
\subsection{\label{sbc:sgs_arch}Overall system architecture}
\begin{figure*}
    \centering
    \includegraphics[width=1\linewidth]{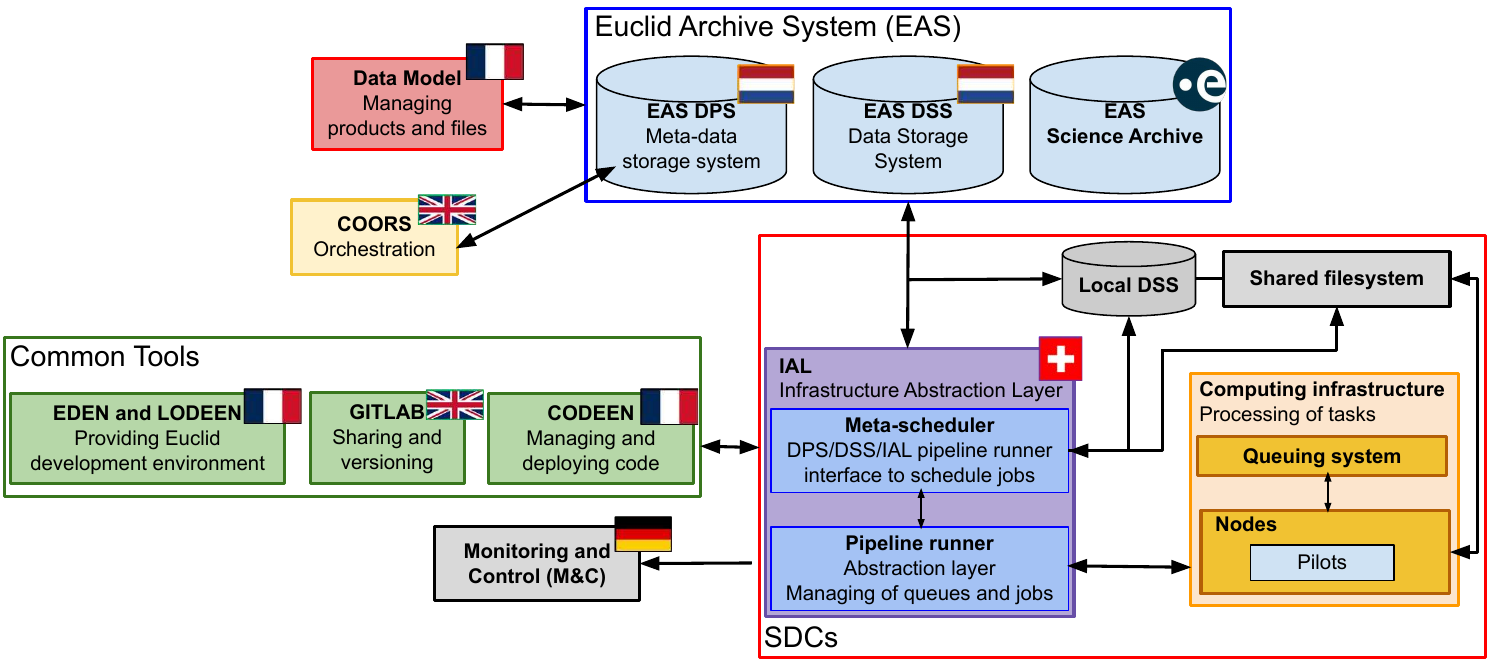}
    \caption{Interaction of system components of the \gls{sgs}.}
    \label{SGS_system}
\end{figure*}

To process the nearly 2500 deg$^2$ of \gls{dr1}, the acquired data are
distributed among the different \glspl{sdc}, each using the full set of pipeline
software developed by the \gls{ou} and \gls{sdc} teams. This distribution limits
inter-cluster transfers and balance the processing load across \glspl{sdc}.
However, ensuring identical processing across different architectures requires
the implementation of a complex system, described below and illustrated in Fig.
\ref{SGS_system}.
A first essential prerequisite for dividing processing by sky portion is the
establishment of a continuous integration and development pipeline. This ensures
that at any given time, a program executed in one \gls{sdc} is identical to that
in another \gls{sdc}. To achieve this, various common resources are made
available to the \glspl{ou} by the System Team (Sect.~\ref{sbc:comm_res}). A
second prerequisite is the implementation of an archive system, referred to as
the \gls{eas}, which facilitates data distribution among the \glspl{sdc} for
processing purposes and the registration of the distributed processing. This
system additionally provides access to scientists for data analysis. The
\gls{eas} is detailed in Sect.~\ref{sbc:eas} with its different components,
namely the \gls{cdm}, the \gls{dps}, the \gls{dss}, and the Science Archive.
Finally, a third prerequisite is to establish an abstraction and orchestration
layer that allows processing to be orchestrated and monitored regardless of the
\gls{sdc} to which it is assigned (Sect.~\ref{sbc:proc_infra}).

\subsection{\label{sbc:comm_res}Common resources}

\subsubsection{\label{ssbc:mdb}Mission Database}

The Euclid Mission Database is a special data product of the \gls{eas}
that stores at any time the official key parameters of the \Euclid mission:
spacecraft and payload parameters, mathematical and physical constants, 
and reference models and calibration products required for data processing by 
the \gls{sgs}, such as the Galactic extinction map and the telescope optical
properties. It is maintained by the System Team, and its updates are managed 
by a Configuration Control Board to ensure that only  approved and validated 
items are included.

\subsubsection{\label{sbc:codeen}Common development and integration environment}
The \gls{ec} \gls{sgs} provides a reference software environment designed to
enforce coding standards and best practices while supporting automated build and
deployment processes. This environment is made available to the \gls{ec} through
a set of common development and integration tools \citep{Poncet2018}.

Source code is hosted on a dedicated GitLab instance, which enables
collaborative development and supports the Git workflow adopted by all \Euclid
projects. Software development relies on the \gls{eden}, which defines the
common standards, tools, and software versions used across the project.
\gls{eden} is implemented as a Conda environment comprising a collection of
packages that install the required software and dependencies\footnote{\url{https://conda-forge.org}}. 
The continuous integration infrastructure, \glspl{sdc}, and the local development 
environments rely on \gls{eden} to build and run \Euclid software.

The codebase is organised using the \texttt{Elements} framework, a C++ and
Python build framework that structures the software into modules grouped into
projects. All \glspl{pf} are implemented within this framework. Continuous
integration is provided by the \Euclid integration platform \gls{codeen}, which
retrieves source code from GitLab and executes a Jenkins \citep{Smart2011}
pipeline that verifies dependencies, builds the project, runs unit tests,
generates and publishes documentation, performs quality analysis, and produces a
SonarQube \citep{Quezada2017} report. The resulting packages are deployed to the
\gls{cvmfs}, which provides a scalable and reliable software distribution
service for the \gls{sgs} \citep{cernvmfs_2020}. Finally, developers are
provided with a \gls{lodeen}, distributed as either a virtual machine or a
Docker image implementing \gls{eden}.

\subsection{\label{sbc:eas}Euclid Archive System}
\subsubsection{\label{ssbc:intro}Introduction} 
The \gls{eas} lies at the core of the \Euclid data-handling system. It registers
all data products, including \gls{le1} outputs, all the \gls{le2} products and
relations (lineage) generated through data processing and archiving, and the
outputs  of the \gls{le3} processing and simulations. The \gls{eas} comprises
three main components: the \gls{dss}, the \gls{dps}, which is built on top of a
central Oracle Database, and the Science Archive. The latter receives the
validated data products produced by the \gls{sgs} for long-term preservation and
scientific access.

The \gls{eas} requires a distributed and scalable architecture, with storage
resources deployed on heterogeneous hardware and file servers across nine data
centres, together with distributed computing capabilities spanning the same data
centres and diverse hardware and software environments. It must also record the
inputs and outputs of the processing pipelines, including file locations and
copies through the \gls{dss}, as well as the parameter values used through the
\gls{dps}, and keep track of all processing and calibration steps together with
the corresponding software versions, thereby ensuring the reproducibility of
individual data products. Finally, it must follow the FAIR principles (Findable,
Accessible, Interoperable, Reusable), with the notion of reusability interpreted
here in the context of reproducibility.

These requirements are met by a data-centric approach based on AstroWise
\citep{begeman2013}, in which a central Oracle database acts as the single
reference for all data products, while the files remain distributed across the
data centres. The common data model is stored in the database through an
object-to-relational mapping: objects defined in the schema are persisted using
the relational model, while their inheritance and association structures are
preserved. This enables full inheritance across all data items, providing
end-to-end data lineage from final products back to every input, including
calibration steps.
The various data releases are delivered by the Science Archive,
which contains a copy of the final validated products of the \gls{eas}, but not the
intermediate products or the data lineage.
In the following sub-sections the various components of the EAS and the Science
Archive are described.

\subsubsection{\label{ssbc:cdm}Common data model} 
A common data model provides a necessary backbone to optimally fullfil the
requirements of the \gls{sgs} data handling. Moreover, the \gls{ec} \gls{sgs}
integrates heterogeneous \glspl{sdc} by settling  common system components in
order to execute a complex processing chain comprising more than 100 pipelines.
To ensure interoperability between these components, the \Euclid \gls{cdm}  also
specifies and controls the interfaces across the \gls{ec} \gls{sgs}.

The \gls{cdm} describes the \Euclid scientific data products generated by the
\glspl{pf}, the \gls{ec} \gls{sgs} infrastructure services and their processing
states, as well as common types and elements shared across several products in
order to avoid duplication. It also defines the structure of the scientific
products stored in the archive as \gls{fits} files.

The \gls{cdm} is implemented as a single \gls{xml} schema describing all
interfaces. Within the processing pipelines, data products are represented as
\gls{xml} files, while the associated data files are primarily stored in the
\gls{fits} format. The use of \gls{xml} scheme ensures that the products
generated by the pipelines comply with the \gls{cdm} specification. In addition,
the \gls{fits} Data Model is defined through a set of \gls{xml} schemas
describing the structure of each \gls{fits} file produced by the pipelines. The
\gls{fits} Data Model and the \gls{cdm} are also used to define the structure of
the \gls{dpdd}\footnote{\url{https://euclid.esac.esa.int/dr/dr1/dpdd/}}, which
provides a comprehensive catalogue of all data products created and processed
within the \gls{ec} \gls{sgs} pipelines, together with their associated metadata
and descriptions.

\subsubsection{\label{ssbc:dps}Data Processing System}
The \gls{dps} implements a data-centric approach to the \gls{eas}. This enables
\gls{sgs} pipelines to use metadata and data files stored in the \gls{eas} as
input and to write resulting data products and files back to \gls{eas}, ensures
reproducibility of generated data items, and allows full traceability of data
product lineage throughout the processing chain. The \gls{dps} serves as the
sole interface between \glspl{pf}. It generates \glspl{ppo} that define
processing tasks, manages the storage and retrieval of data products for
\glspl{pf}, and supports data quality assessment. A data product consists of
metadata represented as an XML file constructed according to the \gls{cdm},
containing references to data files stored in the \gls{dss}.

The core component of the \gls{dps} is a metadata database hosted on an Oracle
RDBMS (Oracle Relational Database Management System), with a schema generated
from the \Euclid \gls{cdm}. The \gls{dps} software layer, named Metadata Access
Layer, implements an object-oriented data model that maps object content to a
relational database. \gls{dps} exposes a \gls{rest api} to the \gls{ial} and 
other \gls{sgs} subsystems \citep{fielding2000}, as well as
to authorised external users. It also provides a graphical interface for
\gls{sgs} users, enabling them to browse data products, manually invalidate
them, flag them for deletion, or trace their lineage.

The operational \gls{dps} instance is provided by SDC-NL and operated at
\gls{esac}. SDC-NL additionally maintains a separate \gls{dps} instance for
testing purposes, covering both data model validation and testing of
\glspl{pf} under development. In both operational and test environments,
\gls{dps} services can be scaled horizontally by deploying additional service
instances as needed. \gls{dps} updates are triggered either by the introduction
of a new \Euclid \gls{cdm} version or by \gls{sgs} requests for new \gls{dps}
functionalities.

\subsubsection{\label{ssbc:dss}Distributed Storage System}
Both \gls{dps} services and \gls{dss} servers use \gls{https} for all
operations. A \gls{dss} server is a thin interface layer on top of the storage
for data files deployed by each \gls{sdc}. The \gls{dss} has a single
centralised metadata database for the registration of file metadata, which
includes file name, file size, checksum, and its location within the \gls{sgs}.
Users can store, retrieve, or delete data files provided they have the
appropriate authorisation and privileges.

\glspl{sdc} can deploy multiple \gls{dss} servers; typically, each \gls{ial} is
accompanied by a \gls{dss} server, with the exception of \gls{esac}, which has
one \gls{dss} server to handle requests from \gls{sgs} users and a separate one
for the ingestion of \gls{le1} data from the \gls{soc} immediately upon
reception. The \gls{ial} is the primary user of the \gls{dss}, since it
retrieves data files for processing and stores newly created data files.

\gls{dss} enforces uniqueness of data file names. In particular, no two files
with the same name but different content can exist in the system. However, it
allows multiple copies of the same file to be created across the network of
\gls{dss} servers. Each \gls{dss} server continuously monitors its connections
to all other \gls{dss} servers in the system and optimises the transfer of
requested data files from the \gls{sdc} with the best connection available at
that moment.

\subsubsection{\label{ssbc:sas}Science Archive}
The Euclid Science Archive, described in \cite{deTeodoro2016} has been designed
to curate the validated data produced by the \gls{sgs} for the scientific
community. In particular, it provides organised access to calibrated and stacked
images of VIS and NIR, MER mosaics in all bands, SIR spectra as well as
catalogues of MER (\Euclid primary catalogue), PHZ (photo-\texttt{z}), and SPE
(spectrocopy) in several tables.  The Science Archive supports standard Virtual
Observatory protocols, ensuring interoperability with external tools and
archives. In particular, it provides a TAP+ (Table Access Protocol) service
enabling Astronomical Data Query Language-based queries (ADQL) on large
catalogues. TAP+ is an extension \citep{Salgado2017} of the IVOA TAP standard
\citep{Dowler2019}. Programmatic access is available through a module within
Astroquery \citep{2019AJ....157...98G}, allowing automated queries, metadata
inspection, and bulk data retrieval in Python workflows. 

The Euclid Science Archive therefore provides the data discovery and retrieval
layer for the validated products of a given data release. These archive services
are integrated with visualisation and scientific analysis capabilities through
the Euclid Data Space, described in Sect.~\ref{sbc:eds}.

\subsection{\label{sbc:eds}Euclid Data Space} 

The Euclid Data Space\footnote{\url{https://euclid.dataspace.esa.int/}}
provides the user-facing environment for accessing and exploiting the scientific
data products delivered by the \Euclid \gls{sgs}. While the \gls{eas} supports
data management and traceability throughout the processing system, and the
Science Archive publishes the final validated products, the Euclid Data Space
brings together data discovery, visualisation, and analysis capabilities through
a common entry point. It therefore provides the interface between the products
generated by the \gls{sgs} and their scientific analysis by the \Euclid
community. 

The Euclid Data Space integrates three main ESA services: the Euclid Science
Archive, ESASky, and ESA Datalabs. The Science Archive provides the
data-discovery layer, including catalogue access through Virtual Observatory
protocols and ADQL queries, as described in Sect.~\ref{ssbc:sas}. ESASky enables
interactive visualisation of \Euclid observations and their exploration in the
context of other astronomical data sets. ESA Datalabs provides the computational
environment in which users can execute software and Jupyter notebooks close to
the \Euclid data. Together, these services support workflows spanning data
discovery, visualisation, and programmatic analysis while reducing the need to
transfer large data volumes to local computing facilities. 

The integration of the Science Archive and ESA Datalabs is particularly
important given the volume of data produced by \Euclid (e.g. 1.7 PB for
DR1), allowing scientific workflows to operate on archive data within the ESA
infrastructure. Programmatic archive access, including through Astroquery, can
be combined with direct access to data files from analysis applications.
Persistent personal workspaces and shared workspaces support individual and
collaborative analyses, while the analysis environments provide commonly used
astronomical packages and mission-specific tools. For \gls{dr1}, these
capabilities are complemented by documentation and Jupyter notebook examples
covering common workflows such as catalogue queries, access to images and
spectra, production of cutouts, and subsequent scientific analysis. 

The public release of \gls{dr1} also marks the launch of the Euclid Data Space
to the wider astronomical community, completing the end-to-end \Euclid data flow
described in this paper: the \gls{sgs} transforms raw observations into
calibrated and validated scientific products, the \gls{eas} preserves their
metadata and processing provenance, and the Science Archive publishes the
validated products of a data release, which can then be discovered, visualised,
accessed, and analysed through the integrated services provided by the Euclid
Data Space.

\subsection{\label{sbc:proc_infra}Processing orchestration and monitoring}
\subsubsection{\label{ssbc:coors}Common Orchestration System}
The \gls{ec} \gls{sgs} data processing is orchestrated through Processing Plans,
which are created and managed using \gls{coors}. It is a web service accessible
through both a \gls{gui} and a \gls{rest api}, allowing users to create and
manipulate Processing Plan objects in the \gls{eas}. A Processing Plan specifies
the processing to be performed by defining a set of parameters, including the
\gls{pf} to be executed, the required calibration and input data, the
observation dates and sky area, and other relevant processing metadata. Based on
these parameters, \gls{coors} invokes a service provided by the \gls{dps} to
expand the Processing Plan into a set of \glspl{ppo}, each defining an
individual processing task. The resulting \glspl{ppo} can either be explicitly
assigned to an \gls{sdc} by the \gls{coors} user or assigned automatically
according to the distribution of the selected sky area across the \glspl{sdc}.
At each \gls{sdc}, an instance of the \gls{ial}, described in
Sect.~\ref{ssbc:ial}, periodically queries the \gls{eas} for \glspl{ppo}
assigned to the local \gls{sdc}. This polling mechanism allows the \gls{ial} to
retrieve newly available processing tasks, schedule them on the local computing
infrastructure, and manage their execution. The resulting products and
processing metadata are registered in the \gls{eas}, providing the link between
the Processing Plan, the individual \glspl{ppo}, and the generated data
products.

Processing plans can also be scheduled based on triggering rules. They can be
turned on and off, allowing for automatic creation of \glspl{ppo} when
new/reprocessed data are available. \gls{coors} also includes functions that
allow monitoring and control of the progress of ongoing processing tasks. Among
others, controls include abortion and restart, software version, and target
\gls{sdc} change. 

\gls{coors} is available through two distinct instances: one connected to the
test \gls{dps} environment and development \glspl{sdc}; and another linked to
the production system. This dual-instance architecture supports both development
activities and operational processing needs.

\subsubsection{\label{ssbc:ial}Infrastructure Abstraction Layer}
The \gls{ec} \gls{sgs} does not impose technological choices on the \glspl{sdc},
allowing each \gls{sdc} to use its own Distributed Resource
Manager and shared file system. The
\gls{ial} acts as the interface between each computing cluster and the rest of
the \gls{ec} \gls{sgs} system, both within and outside the cluster. At the same
time, the \gls{ial} also provides the interface between the \glspl{pf} and the
\gls{ec} \gls{sgs}. \glspl{pf} implement processing pipelines using the Pipeline
Framework provided by the \gls{ial}. The Pipeline Framework enables \gls{pf}
developers to define the processing logic as a data-flow graph.

Two subsystems, namely the Meta-Scheduler and the Pipeline-Runner, provide these
services. The Meta-Scheduler is implemented as a state machine. It retrieves the
\glspl{ppo} assigned to its \gls{sdc} and the associated data products by
communicating with the archive \gls{dps} (Sect.~\ref{ssbc:dps}). It then
prepares the \gls{ppo} working directory by interacting with the local \gls{dss}
(Sect.~\ref{ssbc:dss}) to fetch all required input files. The execution of the
corresponding \gls{pf} pipeline on the cluster is delegated to the Pipeline-Runner. 
After computation, the Meta-Scheduler performs the reverse sequence: it
stores the results in the local \gls{dss}, uploads the newly generated data
products to the archive, and finally sets the \gls{ppo} to its terminal state.
The Pipeline-Runner executes the \gls{pf} pipelines defined as data-flow graphs
in its Python framework. The execution operates on the input data prepared by
the Meta-Scheduler. The graph consists of pipeline steps; in \gls{dr1}, a single
pipeline can comprise up to $10^{6}$ steps. A pipeline step is an executable --
also referred to as a task or job -- that follows strict rules: all inputs and
outputs are file-based; input files are immutable; and each executable declares
its required computing resources. The data-flow graph is evaluated dynamically,
enabling data-parallel processing approaches on generated data (e.g., per galaxy
in a catalogue).

Pipeline steps are not executed directly on the cluster. The Pipeline-Runner
uses the pilot model \citep{casajus2010dirac} to delay scheduling decisions,
prepare execution environments, and profile executables. Pilots are long-running
jobs submitted to the Distributed Resource Manager that can execute multiple
\gls{pf} tasks, thereby reducing queue waiting times. Both pilot generation and
task scheduling rely on a scoring algorithm that considers compatibility between
pilot specifications (defined at the \gls{sdc} level), task requirements, task
priorities, fairness between \glspl{pf}, waiting times, and I/O intensity. The
selected pilot jobs are then submitted to the local Distributed Resource Manager
for execution on available cluster nodes, and \gls{pf} tasks are scheduled to
these pilots for execution. 

The Meta-Scheduler and the Pipeline-Runner provide a local web interface to
monitor and control the execution of \glspl{ppo}. They also transmit telemetry
and profiling data (for each \gls{ppo}, pilot, node, and the \gls{ial} itself)
to the Monitoring and Control subsystem (Sect.~\ref{sbc:monit}), enabling near
real-time monitoring of the complete \gls{ec} \gls{sgs} down to the level of a
single executable. Additional details on the internal design and features of the
IAL can be found in Appendix \ref{app:ial}.

\subsubsection{\label{sbc:monit}Monitoring and control}
To effectively monitor the distributed data processing system, different types
of data are collected directly at the \gls{sdc} using the monitoring ecosystem
built around the \texttt{Elastic
Stack}\footnote{\url{https://www.elastic.co/elastic-stack}} -- a family of tools
for centralized log and metrics management \citep{chhajed2015}. As part of this
ecosystem,
\texttt{Filebeat}\footnote{\url{https://www.elastic.co/beats/filebeat/}}, a
lightweight logs collector, gathers profiling data from the \gls{ial} and
\gls{dss} systems at each node to capture processing states
\citep{CalderonEtal2024}. All of this information is centralised and processed
at SDC-DE. \texttt{Logstash} acts as a processing pipeline, ingesting raw logs
from multiple sources before filtering, parsing, and enriching them. The
filtered data are then stored in \texttt{Elasticsearch}, a search engine that
indexes the documents for fast retrieval and aggregation. Finally,
\texttt{Kibana} provides a web-based interface to explore the indexed data
through interactive dashboards and visualisations.

Once processed, the data are made accessible through a \texttt{Grafana} server
\citep{chakraborty2021}, an open-source platform for interactive dashboards and
time-series visualisation. The dashboards offer a top-down view, where operators
can monitor the overall productions and computing resource usage at all
\glspl{sdc} at a glance. The \gls{sdc} administrators get a more detailed view
of the processing activities at their site, with a plot showing the current
system usage. \gls{pf} developers can drill even further down and analyse the
resource usage of their software while it is running at a remote \gls{sdc}. In case
 errors occur in the data processing, the collected log files can be
retrieved and analysed rapidly by the operations team or pipeline developers in
order to resolve issues in a timely manner. 

\section{\label{sc:processing}SGS Data processing}
\subsection{\label{sbc:proccases}Processing cases}
Depending on the specific processing requirements, the pipeline workflow may
follow different paths. For \Euclid data, two main scenarios are considered,
namely \gls{otf} processing and reprocessing. \gls{otf} processing is designed
to automatically process incoming data in near real time, primarily to monitor
the health of the instruments and to validate newly developed versions of the
\glspl{pf}. Reprocessing, on the other hand, is performed in the context of data
release production, with dedicated regression campaigns carried out beforehand
to verify the correct behaviour of the pipelines.

For each data release, all data collected over a defined period are processed
using fixed versions of the \glspl{pf} and calibration products, which have been
previously validated and delivered by the respective \glspl{ou}. Prior to full
production, a representative subset of the data release is processed as part of
a regression campaign to ensure that each \gls{pf} produces the correct results.
Depending on the outcome, modifications may be made to the code, resulting in
updated \gls{pf} versions for the final production. The validated configuration
is then used to process the complete data set.

\subsection{\label{sbc:procmap}Distribution principles}
\Euclid processing requires manipulating a large amount of data at the nine
\glspl{sdc} of the \gls{ec} \gls{sgs}. In order to efficiently utilise the full
computing power of the \gls{ec} \gls{sgs}, the data are distributed between the
\glspl{sdc} as needed to meet the different processing use cases. Each \gls{sdc}
has a target share of processing of the overall number of observations, as
illustrated in Fig. \ref{SDC_ressources} and described in Sect.
\ref{sbc:sdc_dev}.

For \gls{otf} processing, sky sharing definition relies on two main goals. The
first is the execution of calibration activities for PHZ, VIS, SIR, NIR, and SHE
within the \gls{sdc} responsible for the associated \gls{pf}
(Sect.~\ref{sbc:sdc_dev}). The second is the minimization of downtime during
which a given \gls{sdc} receives no data. To do so, the operations team analyses
the \gls{oss}, provided by \gls{soc}, which contains all observations since the
start of the survey until approximately one year after the date of generation.
From the \gls{oss}, the calibration fields are statically assigned to the
associated primary \gls{sdc} and the \gls{ews} observations are distributed
among the nine \glspl{sdc} according to their respective overall share. In order
to minimise the idle time of the \glspl{sdc}, the observations are split per day
and the assignment cycles through the list of \glspl{sdc} by alphabetical order,
respecting the overall share.

This \gls{otf} strategy allows us to use all \glspl{sdc} on a regular basis, but
increases the share of observations that need to be copied to several
\glspl{sdc} to allow for the full processing (to up to 75\% duplicate
transfers).

For the \gls{dr1} campaign, the strategy had to be changed to optimise
utilisation of the computing resources. The operations team defined the list of
observations that should be part of \gls{dr1} from a technical note describing
the perimeter of the \gls{dr1} provided by the Euclid Science Team. From this
list, the auxiliary and calibration fields were assigned to the \glspl{sdc} with
the smallest share. Then, the \gls{eds} observations were statically assigned to
three different \glspl{sdc}: DE, FI, and NL. Finally, the \gls{ews} observations
were distributed to reach the overall share commitments of each \gls{sdc}.
Contrary to the \gls{otf} case, the observations are globally shared, allowing
us to assign large parts of the sky to a given \gls{sdc}, minimising data
transfer: only 10\% of the observations needed to be duplicated to another
\gls{sdc} to enable processing. A quantitative breakdown of how the roughly 2450
deg² of EWS observations are apportioned among the production SDCs is provided
in Table \ref{tab:EWS_SDC_distribution}. 

\vspace{1em}
\begin{table}
\begin{center}
\setlength{\tabcolsep}{8pt}
\caption{Distribution of processed EWS data across SDCs.}
\label{tab:EWS_SDC_distribution}
\begin{tabular}{lrrr}
\hline\hline
SDC & Observations & deg$^2$ & Share \\
\hline
DE & 148 & 74 & 3.02\% \\
FI & 188 & 94 & 3.84\% \\
FR & 1786 & 893 & 36.47\% \\
IT & 1772 & 886 & 36.18\% \\
NL & 231 & 115 & 4.72\% \\
UK & 772 & 386 & 15.76\% \\
\hline
Total & 4897 & 2448 & 100\% \\
\hline
\end{tabular}
\end{center}
\end{table}

\subsection{\label{sbc:procbudget}Processing budget}
For the production of \gls{dr1}, 6942 \Euclid observations were processed. This
represents 18\,TB of raw frames data for VIS, and 6\,TB of raw frames data for
NISP. Indeed, one VIS raw frame represents 1.2\,GB of uncompressed data, and 400\,MB
compressed, so 2.4\,GB of compressed data at the input per observation.
Similarly, one NISP raw frame represents 200\,MB of uncompressed data, and around
50\,MB compressed, so 800\,MB at the input of NIR and SIR (one \gls{ppo} per band
for NIR, and one \gls{ppo} for SIR). Then, these data follow the workflow of the
\gls{sgs} pipeline. Computing times and output data volume can be found in Table
\ref{tab:DR1_stats} for \gls{le2} \glspl{pf}.

Concerning EXT data, different surveys are used to cover all regions and to have
all required bands \citep{DR1-TP006}. Before
the processing of EXT data, Single Epoch Frames are ingested in the \gls{dss} for each
instrument and each filter. For \gls{dr1}, uncompressed Single Epoch Frames represent
around 315\,GB in terms of storage. These data are then used as input in the EXT
pipeline to produce stacked frames.

\begin{table*}
    \centering
    \caption{Computing resources and output data volume per \gls{pf} for the \gls{dr1} production.}
    \label{tab:DR1_stats}
    \begin{tabular}{
        l
        S[table-format=4.1]
        S[table-format=2.1]
        S[table-format=2.2]
        S[table-format=2.1]
    }
        \hline\hline
        PF &
        \multicolumn{1}{c}{Computing resources} &
        \multicolumn{1}{c}{[\%]} &
        \multicolumn{1}{c}{Output data volume} &
        \multicolumn{1}{c}{[\%]} \\
        &
        \multicolumn{1}{c}{[CPUh/obs.]} &
        &
        \multicolumn{1}{c}{[GB/obs]} &
        \\
        \hline
        VIS                  & 111.6  &  3.2 & 60.00 & 26.0 \\
        NIR                  &  16.7  &  0.5 & 11.00 &  4.7 \\
        MER                  & 334.8  &  9.6 & 59.00 & 25.7 \\
        PHZ                  & 167.4  &  4.8 &  1.80 &  0.8 \\
        SHE PSF fitting      & 372.0  & 10.7 &  0.18 &  0.1 \\
        SHE LensMC and Metacal & 1255.5 & 36.1 &  0.54 &  0.2 \\
        SIR                  & 150.2  &  4.3 & 37.00 & 16.2 \\
        SPE                  & 418.5  & 12.0 &  1.80 &  0.8 \\
        VMPZ                 &   8.4  &  0.2 &  1.80 &  0.8 \\
        EXT -- DES           &  41.9  &  1.2 &  5.00 &  1.9 \\
        EXT -- CFIS          &  27.9  &  0.8 & 22.00 &  9.4 \\
        EXT -- WHIGS         &  27.9  &  0.8 & 15.00 &  6.5 \\
        EXT -- WISHES        &  44.6  &  1.3 & 13.00 &  5.6 \\
        EXT -- Pan-STARRS i  & 502.2  & 14.4 &  3.00 &  1.3 \\
        \hline
        Total & 3479.6 & & 231.1 & \\
        \hline
    \end{tabular}
\end{table*}

Each data release requires reprocessing all previously collected data with the
newer \gls{pf} versions to create homogenous data products. Therefore, the
computing resources grow significantly with each new scheduled data release.
This increase in resource requirements is addressed in the \gls{sgs} roadmap,
advising procurement strategy to the \glspl{sdc} to ensure that it is in line
with the output data volume and the required \gls{cpu} hours needed for a given
data release. The roadmap estimates the number of observations to be processed
per year from the survey and the overall \gls{sgs} release planning. This
information is then combined with the \gls{pf} performance statistics collected
by the \gls{ial} profiling (Appendix \ref{app:SGS technical budget}) to
produce the total compute budget needed to complete a data release. Those
figures are amended with retry factors to cover for crashes, \gls{pf} bugs, and
system issues to compile a final budget. Finally, this budget is allocated
proportionally based on each \gls{sdc}'s share, giving them a clear estimate of
the \gls{cpu} and storage resources required per year. The processing budget is
summarised in Table \ref{tab:SGS_resources_year}.

\begin{table*}[htp!]
    \centering
    \caption{Target roadmap of \gls{sgs} computing and storage requirements (number of cores and total TB).}
    \label{tab:SGS_resources_year}
    \begin{tabular}{llrrrrrrr}
        \hline\hline
        \gls{sdc} & Resource & 2024 & 2025 & 2026 & 2027 & 2028 & 2029 & 2030 \\
        \hline
        SDC-CH & Cores   & 163 & 237 & 154 & 343 & 346 & 799  & 243 \\
               & Storage & 165 & 393 & 377 & 665 & 654 & 1299 & 1550 \\
        \addlinespace

        SDC-DE & Cores   & 375 & 543 & 348 & 786 & 791 & 1828 & 545 \\
               & Storage & 387 & 843 & 810 & 1386 & 1365 & 2656 & 3157 \\
        \addlinespace

        SDC-ES & Cores   & 163 & 237 & 154 & 343 & 346 & 799  & 243 \\
               & Storage & 165 & 393 & 377 & 665 & 654 & 1299 & 1550 \\
        \addlinespace

        SDC-FI & Cores   & 273 & 407 & 254 & 593 & 566 & 1349 & 393 \\
               & Storage & 216 & 444 & 581 & 869 & 858 & 1566 & 1817 \\
        \addlinespace

        SDC-FR & Cores   & 813 & 1183 & 770 & 1715 & 1728 & 3995 & 1213 \\
               & Storage & 825 & 1966 & 1884 & 3323 & 3270 & 6497 & 7749 \\
        \addlinespace

        SDC-IT & Cores   & 813 & 1183 & 770 & 1715 & 1728 & 3995 & 1213 \\
               & Storage & 825 & 1966 & 1884 & 3323 & 3270 & 6497 & 7749 \\
        \addlinespace

        SDC-NL & Cores   & 325 & 473 & 308 & 686 & 691 & 1598 & 485 \\
               & Storage & 330 & 786 & 753 & 1329 & 1308 & 2599 & 3100 \\
        \addlinespace

        SDC-UK & Cores   & 325 & 473 & 308 & 686 & 691 & 1598 & 485 \\
               & Storage & 330 & 786 & 753 & 1329 & 1308 & 2599 & 3100 \\
        \addlinespace

        SDC-US & Cores   & 183 & 257 & 164 & 373 & 376 & 879  & 263 \\
               & Storage & 166 & 394 & 382 & 670 & 660 & 1307 & 1557 \\
        \hline
    \end{tabular}
\end{table*}

While we have real numbers for \gls{dr1}, numbers for DR2 and DR3 are
projections based on what was needed for \gls{dr1} and the number of
observations for the next releases. These figures will change however, since the
\glspl{ou} will deliver new versions of the \glspl{pf} with different
performances from those used in \gls{dr1}.

\section{\label{sc:operations}Operations}
\subsection{\label{sbc}Daily processing and nominal operations}
Science data from the VIS and NISP instruments are produced for each exposure in
a compressed format and stored in instrument-specific files within the on-board
mass memory. The timing for opening and closing these files, and consequently
their content, is defined by the \gls{soc} during the planning phase. The CCSDS
File Delivery Protocol, described in \cite{CCSDS7201G4}, is used to send the
files to the ground, ensuring completeness and supporting automatic
retransmission when necessary. One of three ground stations, Cebreros (Spain),
New Norcia (Australia), or Malargüe (Argentina), is used during the roughly
four-hours daily pass \citep{Rudolph2016}. The pass duration provides sufficient
margin to recover any backlog, and daily data transmission averages
approximately 100\,GB. In addition to science data, spacecraft and instrument
\gls{hktm} are embedded in several of these CCSDS File Delivery Protocol files.

Upon reception, files are sent from the ground station to the \gls{moc}, where
the \gls{hktm} packets are extracted, time-stamped, and stored in \gls{moc}
systems. Science files are forwarded to the \gls{soc} as soon as all files are
acquired, typically within four hours after pass completion. The complete
\gls{hktm} data set is also replicated from \gls{moc} to \gls{soc} via the
\gls{ares}, a dual-population database hosting parametric samples of \gls{hktm}
and other data sources. The \gls{soc} ingests information from \gls{ares} and
planning data, typically completing synchronisation within 12 hours of pass end.
For each science file, the \gls{soc} verifies availability of overlapping
\gls{hktm}; when available, processing begins to generate the \gls{le1} products to
extract exposures, to decompress and to reorder pixel data, and to produce a
metadata file with its associated science pixel data as a \gls{fits} file. Additional
metadata required for processing, including calibration parameters, pointing
information, and instrument settings, are added from \gls{hktm} and planning
data in metadata and FITS headers. The final \gls{le1} products are archived in the
\gls{soc} \gls{dss} and distributed to the \glspl{sdc}. This workflow is fully
automated using parallel Docker containers.

Alongside \gls{le1}, the \gls{soc} extracts hourly \gls{hktm} and archives them
as FITS-format \gls{hktm} products in the \gls{dps}/\gls{dss}. These data sets
support subsequent \gls{ec} \gls{sgs} processing and serve as a safeguard
against potential future reprocessing requirements. The \gls{soc} also performs
additional analyses on \gls{le1} products through the \gls{qla}, which evaluates
exposure consistency and derives values for instrument health monitoring.
\gls{qla} is also applied to \gls{hktm} attitude parameters to detect potential
pointing deviations; these outputs are used internally within the \gls{soc} but
are not archived in the \gls{dps}/\gls{dss}.
\begin{figure}
    \centering
    \includegraphics[width=1\linewidth]{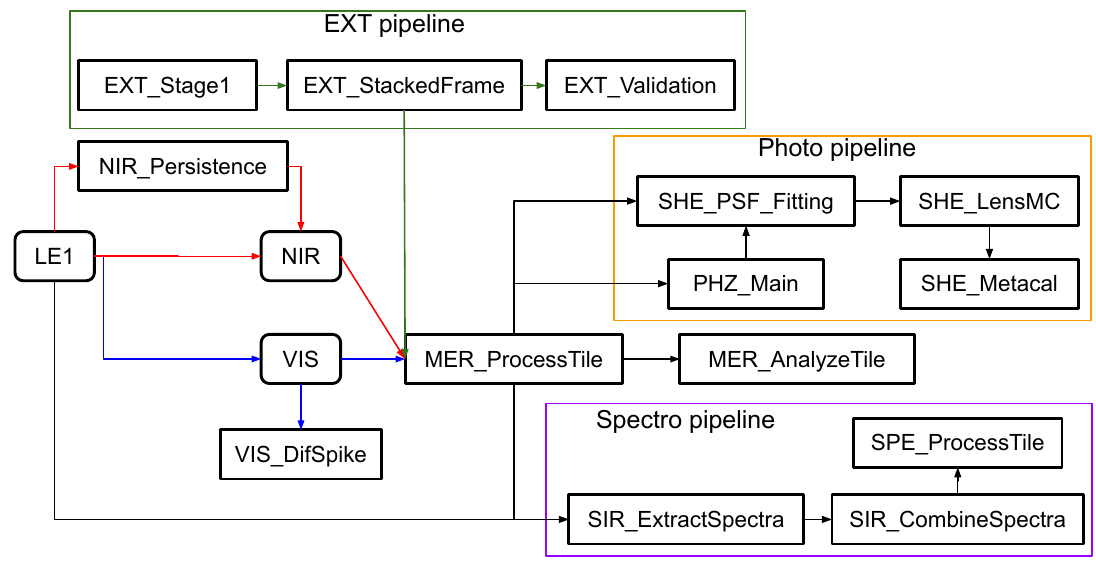}
    \caption{Illustration of the \gls{otf} processing in the \gls{sgs}.}
    \label{Processing_pipeline_SGS}
\end{figure}
The \gls{otf} processing strategy enables automatic execution of \glspl{pf}
immediately after raw observations are ingested and validated within the
\gls{sgs}. Managed by \gls{coors}, this automation reduces operator workload and
accelerates processing by triggering re-execution on newly ingested sources.
\glspl{pf} are categorised by operational scale, as illustrated in
Fig.~\ref{Processing_pipeline_SGS}, to handle the spatial complexity of \Euclid
observations, which range from individual exposures to fixed-grid MER tiles
\citep{Q1-TP004}. Two types of tile have been defined: wide tiles, with a core
area size of $ 30\arcmin \times 30\arcmin$; and deep tiles, with a core area
size of $ 15\arcmin \times 15\arcmin$. Initial reductions, for both VIS and NIR,
operate at the observation scale, while algorithms such as
\texttt{SIR\_ExtractSpectra} and \texttt{SHE\_PSF\_Fitting} also run at this
level. Data integration begins at the tile level with \texttt{MER\_ProcessTile},
aggregating VIS, NIR, and EXT stacked frames, and subsequent \glspl{pf} (PHZ,
SHE, SPE) continue at the tile level to ensure consistency across the survey
grid.

This hierarchical structure introduces complexity in managing \gls{le2}
pipelines triggering. Multiple observations are often required to cover a single
MER tile, acquired over different days, and the system must balance timely
processing with resource efficiency. Triggering decisions are guided by
\texttt{IntersectionMaps} and triggering rules, which track theoretical and
processed relations between observations and tiles, coverage ratios, and
acquisition dates. In particular, \texttt{IntersectionMaps} cross-references
telescope observations with tiles to track area coverage and automatically
trigger data processing pipelines. Tile-level \glspl{pf} such as
\texttt{MER\_ProcessTile} are executed once all expected observations are
processed, a coverage threshold is met, or a defined delay has elapsed, whereas
observation-level \glspl{pf} await availability of sufficient MER catalogues.

In addition to the \gls{ews} and \gls{eds}, calibration observations as defined by
the Euclid Calibration Framework are scheduled via the OSS. Two primary
activities are performed: self-calibration executed monthly with flexibility to
accommodate survey constraints; and VIS PSF monitoring/NISP photometric
nonlinearity checks performed every two to six weeks, generally after
significant thermal disturbances. The operations team uses \gls{oss}-based tools
to identify acquisition dates and launch sequences of \glspl{pf} automatically.
For self-calibration observations, VIS and NIR \glspl{pf} are run in parallel.
For VIS PSF monitoring/NISP photometric nonlinearity checks, VIS and NIR
\glspl{pf} are processed in parallel, followed by MER and PHZ. This ensures
systematic data processing and continuous instrument health monitoring.

Automation relies on Argo Workflows~\citep{Yuen2021}, a cloud-native,
open-source workflow engine that orchestrates containerised tasks as directed
acyclic graphs. In the SGS, Argo Workflow templates define independent
end-to-end processing workflows that orchestrate \gls{pf} execution. The
workflows submit \glspl{ppo} to the \gls{dps}, monitor their execution status,
and trigger subsequent steps only when upstream tasks succeed with the expected
inputs.

\subsection{\label{sbc:iot_activities}Activities of the Instrument Operation Teams}

One of the primary responsibilities of the \gls{iot} is to verify the
operational status and health of the \Euclid NISP and VIS instruments, a task
performed both on a daily and weekly basis. For this purpose, a dedicated tool
called \gls{ioda} has been developed, which is used by both the VIS and NISP
teams \citep{Riccio2022}. By accessing the \gls{hktm} and \gls{qla} data,
\gls{ioda} enables the generation of monitoring reports based on a list of
predefined parameters, accompanied by diagnostic graphs and trend analyses. The
\gls{ares} system is also used by the \glspl{iot} to quickly check on \gls{hktm}
and telecommand data.

In order to monitor the behaviour and trends of the instruments, the \gls{iot}
is also responsible for validating calibration products that are generated
periodically over the survey. The validation process undergoes specific
assignments: the first validation level is the responsibility of the NISP and
VIS instrument scientists who must verify that the instrument performance aligns
with expectations; and the validation is supported by the Calibration Support Team
and Instrument experts, who use additional tools beyond \gls{ioda}. Once the
product has been validated by the Instrument Scientists, the Calibration
Scientist must verify that the calibration observation result is compliant with
the mission scientific requirements. Once this second assignment is confirmed,
the calibration product is declared valid.

Finally, the \gls{iot} handles instrument maintenance: when necessary, the
\gls{iot} sends  Instrument Commanding Requests to be included in the mission
plan with the aim of improving instrument performance or safety. These requests
are prepared and verified using a dedicated software tool, the  Instrument
Commanding Request tool, which converts the set of instructions to be executed
into XML format. The transmission procedure of the request varies depending on
the potential impact of its execution to the survey, and the same applies to the
interfaces (\gls{soc} or \gls{moc}).

\section{\label{sbc:dr1prod}Data Release 1 production} 
\begin{figure}
    \centering
    \includegraphics[width=1\linewidth]{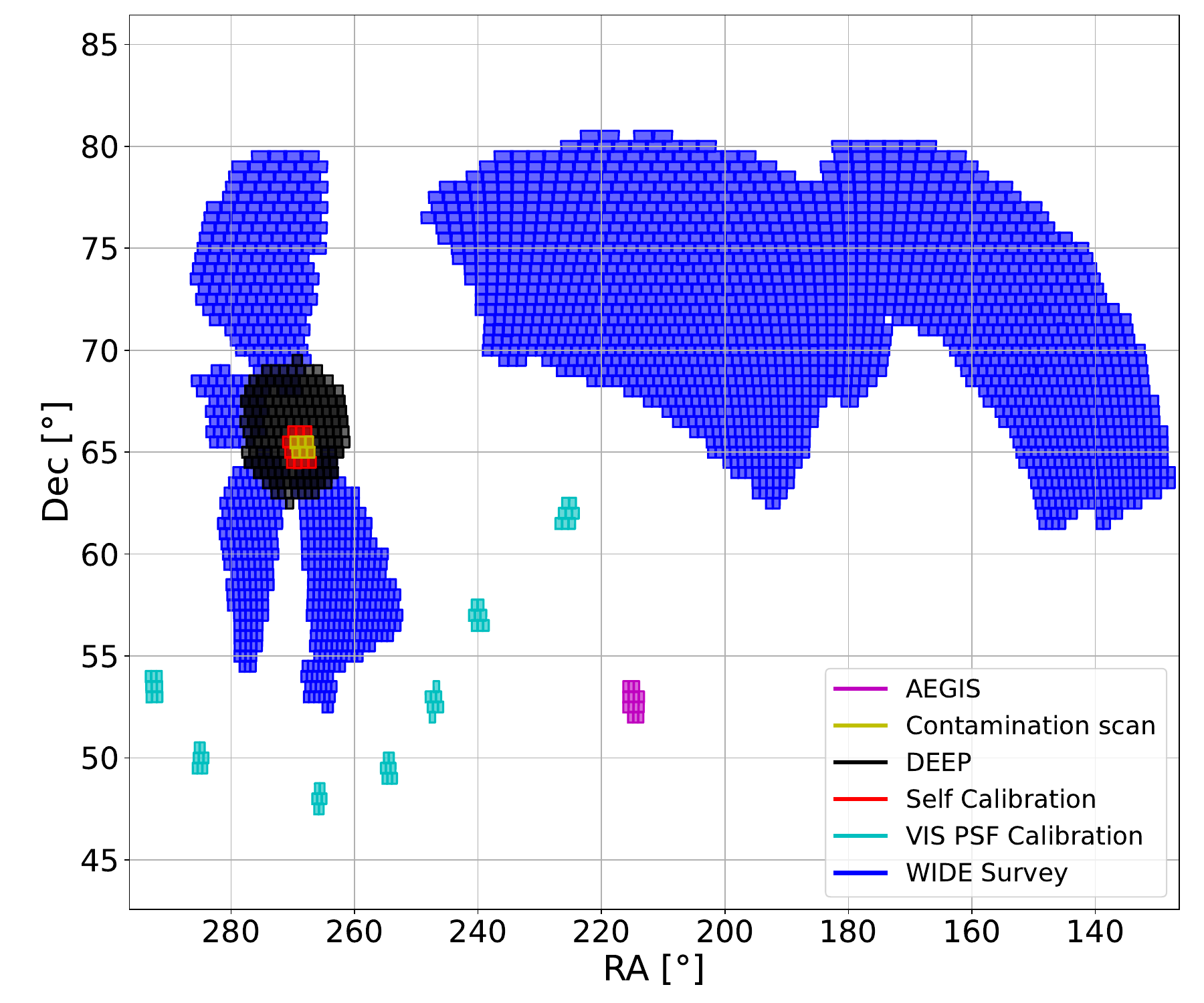}
    \caption{Input northern coverage of \gls{dr1}.}
    \label{fig:north_coverage}
\end{figure}

\begin{figure}
    \centering
    \includegraphics[width=1\linewidth]{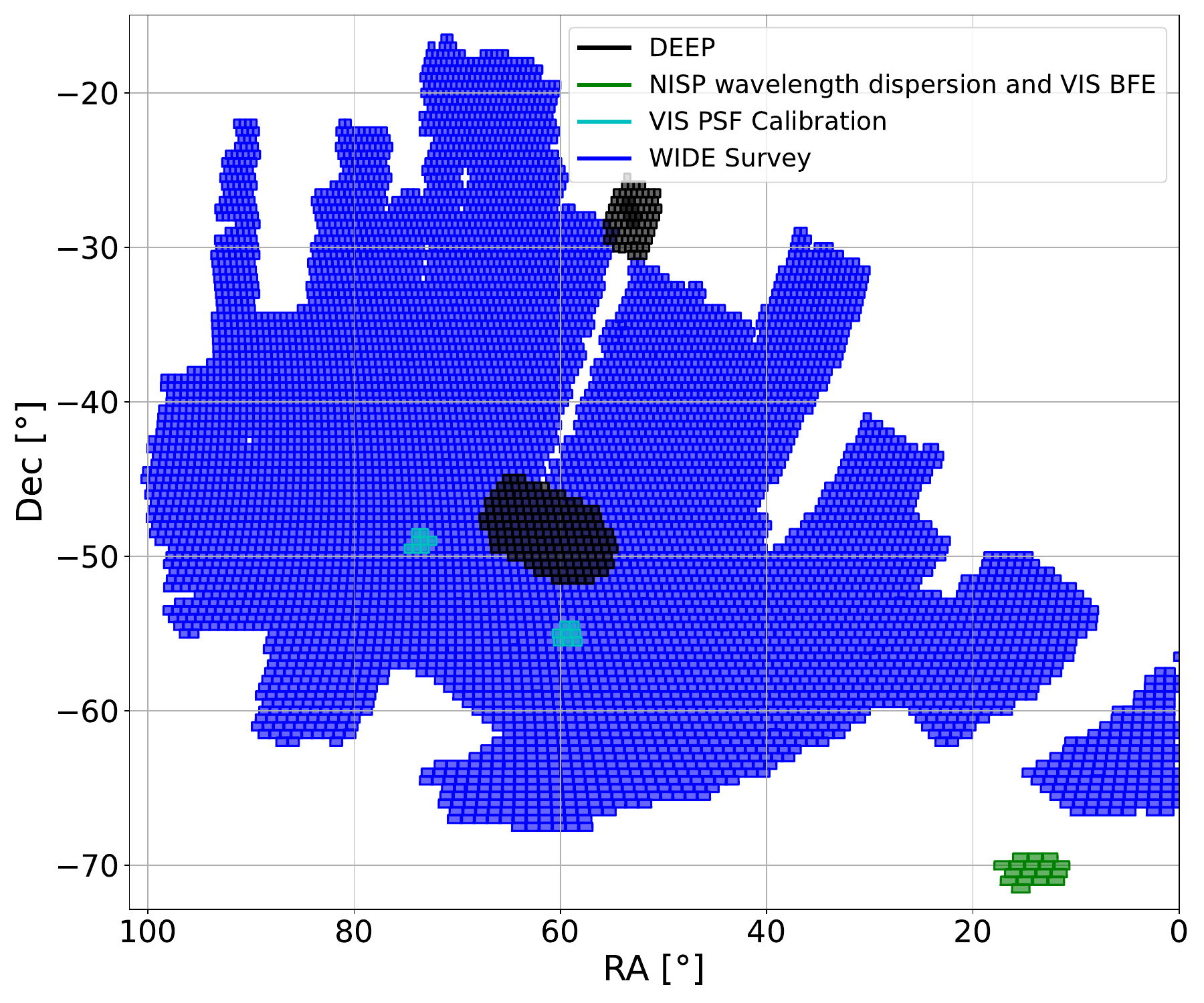}
    \caption{Input southern coverage of \gls{dr1}.}
    \label{fig:south_coverage}
\end{figure}

The survey and the scope of the \gls{dr1}  are described in detail by
\cite{DR1-TP001} and \cite{esa2026}; here we focus on the SGS processing
aspects.  The \gls{dr1} production campaign encompasses \Euclid observations and
calibration data collected between February 2024 and May 2025. This extensive
data set comprises the \gls{ews} fields (2375 deg²), three \gls{eds} fields
(North, South, and Fornax, totalling 53 deg²), and eight specialised categories
covering Performance Verification, Calibration, Auxiliary, and Phase Diversity
Calibration fields. For all target areas, the final observation list excludes
data impacted by instrument decontamination campaigns, technical reboots, or
periods of high solar activity. Additionally, observations intersecting the RA =
0° meridian were omitted from \gls{dr1} due to current geometric constraints in
handling the celestial coordinate wrap-around; these are slated for future
inclusion following algorithmic updates. Beyond these hardware and geometric
filters, the execution of pipelines requiring external ground-based data remains
contingent upon confirmed coverage, availability, and successful completion of
EXT-specific processing.

The workflow begins with the generation of essential calibration products, such
as instrument reference frames and detector characterisation data, which require
formal approval by the \gls{ec} \gls{sgs} Configuration Control Board and
instrument scientists. To manage survey diversity, processing is partitioned by
field type, utilising tailored execution sequences. For the \gls{ews},
production initiates with the parallel execution of the VIS, NIR, and EXT
\glspl{pf}. The VIS and NIR branches process \gls{le1} data and produce
calibrated exposures, incorporating specialised steps such as NIR persistence
masking and VIS diffraction-spike monitoring. These parallel streams are
mandatory prerequisites for the MER \gls{pf}, which detects sources, measures
their properties, and combines them into an unified multi-wavelength catalogue.
Post-MER, the workflow branches into the photometric pipeline with PHZ and SHE,
and the spectrometric pipeline with SIR and SPE, as illustrated in Fig.
\ref{Processing_pipeline_SGS}.

In contrast, \gls{eds} processing follows a two-fold hierarchical workflow.
Initially, each individual visit is reduced independently in a wide-like mode,
with NIR, VIS, and MER processing performed at the wide tile level using
external datasets down-selected to match the \gls{ews} depth. For red-grism
observations, a specific spectroscopic chain is triggered to produce unitary
catalogues. Once validated, a consolidated full-stack mode is initiated during
reprocessing campaigns to integrate all passes. This deep mode leverages
specialised external data sets (e.g., HSC; \cite{EP-McPartland};
\cite{EP-Zalesky}; \cite{Aihara+22}; and DES, \citealt{Abbott+21}) and employs a
dedicated PHZ Deep pipeline, alongside a dual-run spectroscopic combining
process that merges blue and red grism data at the deep tile level.

\begin{figure}
    \centering
    \includegraphics[width=1\linewidth]{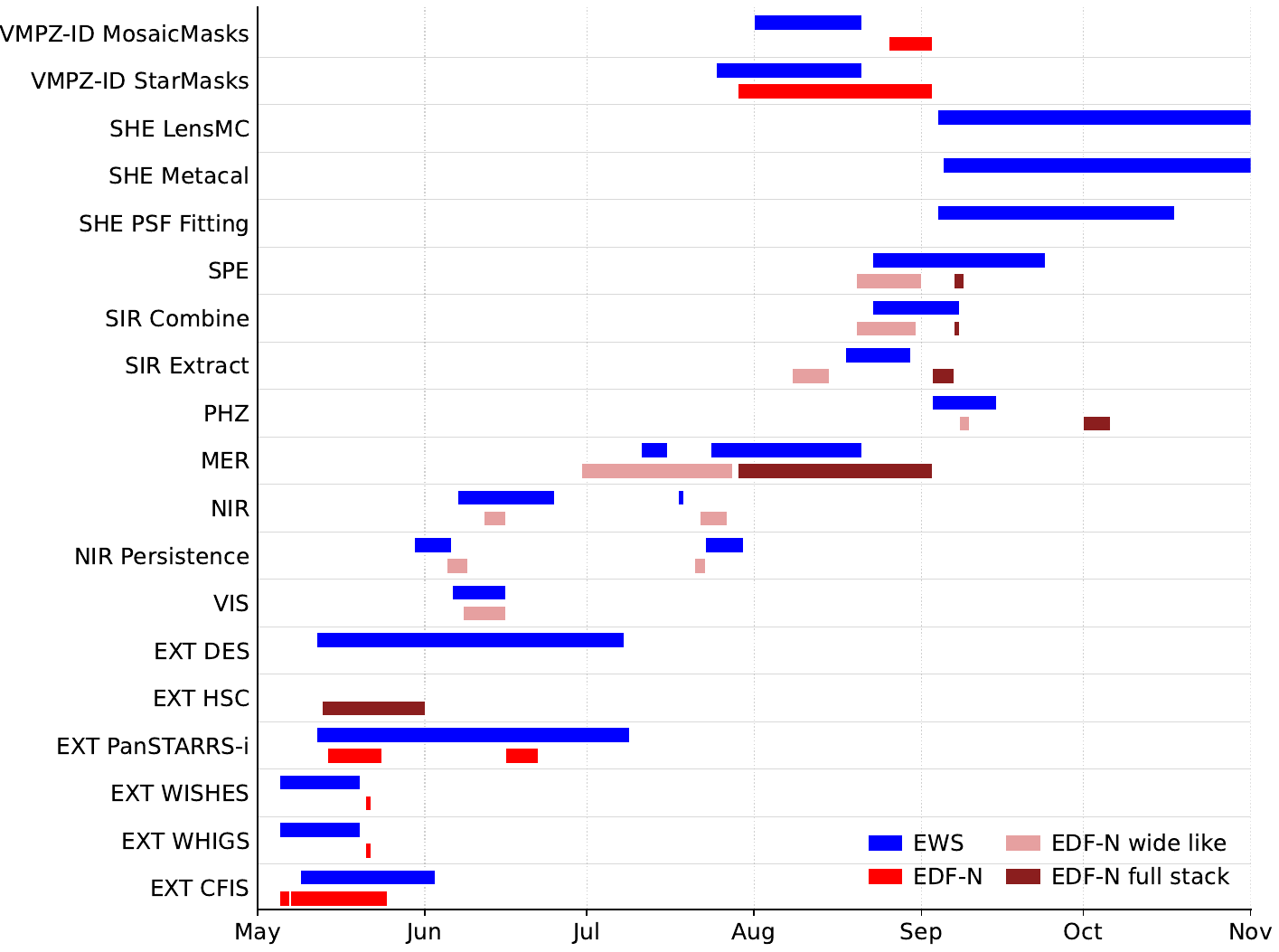}
    \caption{\gls{ews} and \gls{edfn} main processing timelines.}
    \label{DR1_Processing_Timeline}
\end{figure}

The operational timeline was dictated by the staggered availability of these
external data sets and the sequential dependencies of the scientific pipelines.
As illustrated in the \gls{ews} and the \gls{edfn} main processing timelines in
Fig. \ref{DR1_Processing_Timeline}, the campaign was launched in May 2025 with
the processing of EXT data sets (DES, Pan-STARRS, WISHES, WHIGS, and CFIS).
Processing of other fields, including the \gls{edfs}, \gls{edff}, and COSMOS,
began at the same time. Additional processing steps were performed after these
main sequences to recompute specific fields for the PHZ and SHE products, as
well as for the subsequent SIM and LE3 processing stages.

These preliminary steps proved operationally intensive due to multiple
redeliveries and technical challenges involving high-volume data. Parallel
processing for VIS and NIR began in June 2025, leading into the MER phase.
Unlike the \gls{otf} processing, MER has not been launched automatically, but
only after VIS and NIR data validation by the corresponding \gls{ou}. However, a
calibration error identified during validation necessitated the reprocessing of
13 sky patches in the \gls{ews} and seven in the \gls{eds}. This impacted MER
results, requiring their invalidation and subsequent relaunch. Downstream
execution for the spectrometric and photometric pipelines followed, utilising
the automated triggering system driven by dedicated \texttt{IntersectionMaps}
and specific business rules. Despite this automation, operational dependencies
remained; for instance, late deliveries of PHZ shifted the subsequent
commencement of SHE processing into late September 2025. Versions of \glspl{pf}
used for the \gls{dr1} processing can be found in Appendix \ref{app:version
PFs}.

To maintain scientific integrity, all products undergo validation on a sky-patch
basis via Data Quality Check pipelines. This rigorous monitoring is vital for
early error detection, as demonstrated during \gls{dr1} when specific
calibration issues necessitated the reprocessing of several patches across both
\gls{ews} and \gls{eds} fields. The scale of this mission is reflected in its
significant computational footprint: 34.4 million jobs executed via 200 830
\glspl{ppo}, 774 CPU years (approximately 6.7 million hours), a final archive
size of 1.31\,PB (excluding EXT data), and an operational error rate of only
0.7\% over all \glspl{ppo} executed.

\section{\label{sc:conclusions}Conclusions}
In this paper we have presented the architecture of the \Euclid \gls{sgs}, the
role of the \gls{soc}, the \glspl{sdc}, the \gls{sgs} common infrastructure, and
the processing methodology implemented by the \gls{ec} \gls{sgs} for the
\gls{otf} and also for the production of the Euclid \gls{dr1}. The system
integrates a distributed network of \glspl{sdc}, a modular pipeline
architecture, and a coordinated validation framework designed to process the
large data volumes produced by the \Euclid survey.

The \gls{dr1} production has demonstrated the capability of the \gls{sgs}
to operate at survey scale while maintaining strict requirements on
data quality, traceability, and reproducibility. The pipeline framework enables
the execution of complex processing chains while ensuring consistent
configuration management and versioned data products. From the beginning of the
\Euclid mission, it was acknowledged that handling the data volumes required for
processing poses a challenge in itself. As of January 2026, the total number of
files stored in the \gls{dss} reached 80 million, with a total data volume of
4.5\,PB (distributed across all \glspl{sdc}), while the metadata databases
(\gls{dps} and \gls{dss}) have a moderate volume of 7.5\,TB.

The experience gained during the \gls{dr1} production phase has been instructive
in validating the end-to-end processing system, as well as identifying
performance and validation process bottlenecks. In particular, the coordinated
efforts of the System Team, \gls{pf} developers, \glspl{sdc}, and \glspl{ou} in
the validation process have enabled the consolidation of operational procedures
and tools required for large-scale product validation.

The infrastructure and operational model described here provide the foundation
for the forthcoming \Euclid data releases, which will involve significantly
larger survey areas and increasing data volumes. Continued evolution of the
\glspl{pf}, \gls{sdc} resources, and validation framework will ensure the
long-term stability and efficiency of the \Euclid \gls{sgs} and support the
delivery of high-quality data products to the astronomical community throughout
the mission lifetime.

%
%

\begin{acknowledgements}
\AckEC  
\end{acknowledgements}

%
%

\bibliography{Euclid, DR1, SGS} 
%
%

%

\begin{appendix}

\section{Acronyms}     \vspace{-0.5cm}
\printnoidxglossary[type=\acronymtype, title={}]

\clearpage
\section{\label{sec:le3_pf}\Euclid LE3 analysis pipelines}

The \gls{le3} data analysis pipelines are designed to extract cosmological
information from the survey data products generated by the upstream processing
levels. While the previous processing stages focus on instrument calibration,
image reduction, and catalogue production, the \gls{le3} pipelines operate on
high-level data products to derive measurements of summary statistics that are
used for cosmological inference. \gls{le3} is the first stage of \gls{sgs}
processing in which the full sky area of a given data release is analysed
jointly, and the last stage at which data quality validation is performed. This
is critical, since certain systematic effects only become detectable once a
sufficiently large sky area or galaxy sample has been processed.

The primary cosmological probes targeted by \gls{le3} are \gls{wl} and \gls{gc}.
Each probe employs an independent data-processing strategy that requires careful
selection of target sources from the main catalogue: galaxy shapes for \gls{wl};
and the 3-dimensional distribution of galaxies for \gls{gc}. These samples are
then used to measure specific summary statistics that characterise the
large-scale structure of the Universe, including correlation functions, power
spectra, and higher-order statistics, together with the associated covariance
estimates.

Within the \Euclid \gls{sgs} architecture, the different analysis tasks are
implemented through dedicated \glspl{pf}, which represent the fundamental
execution units of the \gls{le3} pipelines. Individual analysis components are
therefore organised into well-defined projects responsible for specific stages
of the workflow, including sample selection, preparation of analysis-ready
catalogues, and the computation of statistical estimators.

The spectroscopic \gls{gc} analysis is handled by several \glspl{pf}. The first
step of the workflow is to apply the target selection to the MER catalogue, and
creating the \gls{le3} spectroscopic galaxy catalogue. This process involves
selecting sources that satisfy the spectroscopic quality and redshift
reliability criteria, ensuring that the final catalogue contains objects
suitable for cosmological measurements. In addition to generating the data
catalogue, the corresponding random catalogue is produced. This represents a
reference sample of galaxies distributed randomly over the survey volume, and is
required by the downstream \gls{gc} \glspl{pf} that measure clustering
statistics. The random catalogue provides the unclustered baseline against which
the observed galaxy distribution is compared, allowing the statistical
estimators to properly quantify deviations from spatial uniformity. The
construction of the random catalogue is a critical component of the analysis
pipeline. In order to enable an unbiased extraction of cosmological information,
the random sample must closely reproduce the survey geometry, angular footprint,
and observational selection effects present in the data catalogue. This includes
for example, the survey mask, variations in observational completeness, and
other systematic effects affecting the spectroscopic selection.

A specific summary statistic is then computed by providing the data and random
catalogues as input to the corresponding \gls{le3}-\gls{gc} \gls{pf}, together
with a dedicated configuration file that specifies the details of the analysis.
Several clustering statistics can be measured within the \gls{gc} software
suite: the \gls{2-pcf}, the \gls{3-pcf}, the power spectrum, and the bispectrum.
In addition, the numerical covariance matrices associated with the 2-point
statistics are computed. Each \gls{le3}-\gls{gc} \gls{pf} supports multiple
estimator implementations, which can be selected through the configuration file
provided as input to the code. The configuration file also allows the user to
define several analysis parameters, including algorithm-specific settings and
the binning strategy adopted for the measurement of the clustering statistics. A
detailed description of the codes can be found in \citet{DeLaTorreEtal2025} for
the \gls{2-pcf} estimator, whereas for the power spectrum estimator the
corresponding paper is in preparation. In summary, for the power spectrum we
adopt the estimator proposed by \citet{Yamamoto:2006}, but based on fast Fourier
transforms, for an efficient evaluation following \citet{Scoccimarro2015} and
\citet{BianchiEtal2015}.

Each \gls{pf} is executed, within the \Euclid \gls{sgs} infrastructure, by a
dedicated pipeline that automates the workflow and handles data product
retrieval and storage. The current \gls{le3}-\gls{gc} pipelines enable
computation of all  the aforementioned statistics in the nominal configuration,
which takes as input one data catalogue, one random mask, and a parameter file.
It is also possible to compute the 2-point statistics cross-estimators
pipelines feeding, on top of the nominal inputs, another set of data and random
catalogues. Lastly, the 2-point statistics pipelines can also be run using the
estimators for the reconstruction by giving an additional random (reconstructed)
catalogue in addition to the nominal set of galaxy and random catalogues.

The photometric branch of \gls{sgs} processing encompasses two cosmological
probes:  $3{\times}2$-pt, which falls under the umbrella of \gls{wl}; and
\gls{cl}. Both probes receive their input products from VMPZ-ID, which is the
primary interface between \gls{le2} and \gls{le3} products. VMPZ-ID comprises
several pipelines (such as \texttt{CatRed}, \texttt{Mosaic}, \texttt{Star},
\texttt{Galaxy}, and \texttt{Concat}) that collectively produce a range of data
products. In particular, the \gls{pf} provides a set of per MER tile masks
describing instrumental properties (such as depth, noise, point spread function,
and coverage), survey properties (such as the survey footprint), and photometric
sky foregrounds (such as Galactic extinction and zodiacal light). It also
provides visibility masks and random galaxy catalogues for photometric galaxy
clustering, as well as a set of masks representing catalogue properties, reduced
galaxy catalogues, and finally concatenated versions of the previous per tile
products. These reduced galaxy catalogues are the primary inputs for the
\gls{wl} and \gls{cl} pipelines and critically allow for science-specific
selections to be consistently propagated, facilitating the reproducibility of
scientific results.

The \gls{wl} pipelines encompass the full $3{\times}2$-pt data vector, combining
cosmic shear, galaxy-galaxy lensing, and photometric galaxy clustering, measured
in both real (or configuration) space and harmonic space, along with the
corresponding covariance matrices. The \gls{2-pcf} \gls{wl} pipeline is based on
the \texttt{athena} code \citep{kilbinger:2014, schneider:2002} and provides the
real-space correlation functions, band power spectra, and Complete Orthogonal
Sets of $E$-/$B$-mode Integrals \citep[COSEBIs;][]{schneider:2010}, which are a
key test for $B$ modes arising from systematic errors. The pipeline implements
both the \citet{landyszalay:1993} and \citet{hamilton:1993} estimators for the
angular correlation functions, and jackknife and bootstrap resampling for the
covariance matrices. The power spectrum \gls{wl} pipeline implements the
\texttt{Heracles} code \citep{tessore:2025} to produce harmonic-space angular
power spectra and mixing matrices. Additionally, covariance matrices are
produced via the jackknife-based DICES algorithm \citep{Naidoo2026}. In addition
to the 2-point statistics, convergence (or mass) maps based on two different
algorithms are provided. The first method is the standard method proposed by
\citet{kaiser:1993}. The second algorithm employs the nonlinear inversion method
\texttt{KS+}, described in \citet{pires:2020}. Given the large survey area
(nearly 14\,000~deg$^{2}$) mass mapping can be carried out either in the plane by
splitting the survey into small fields, where the flat-sky approximation remains
valid, or directly on the curved sky. \Euclid will provide convergence maps for
the entire survey area using both strategies, along with high-resolution maps of
smaller patches centered on the most massive clusters of galaxies.

VMPZ-ID products are also used by the \gls{cl} pipelines. The \gls{cl} pipelines
are designed to provide a catalogue of galaxy clusters and to characterise their
properties. The first step is the detection of clusters via two cluster finders:
an Adaptive Matched Indentifier of Clustered Objects \texttt{AMICO}
\citep{Bellagamba2019}; and a wavelet-based algorithm \texttt{PZWAV}
\citep{Werner2023, Thongkham2024}. They have been selected through the final run
of Cluster Finding Challenges, in which the performances of six cluster finders
have been compared \citep{Adam2019}. Then, the selection function is estimated
for each cluster catalog resulting from \texttt{AMICO} and \texttt{PZWAV} using
a data-driven approach \citep{Maturi2019}. The main properties of cluster
candidates are then calculated: the cluster mean spectroscopic redshift and its
velocity dispersion when sufficient spectroscopic data are available, as well as
the richness and the weak lensing masses. Two estimates of the richness of the
cluster candidates are provided based on photometric redshift estimates
\citep{CastignaniBenoist2016} and on the red sequence properties of cluster
galaxies \citep{Andreon2016}, as well as, for each galaxy, the membership
probability of belonging to a given cluster candidate.  Various distributions
are also estimated, such as  weak lensing shear profiles and semi-analytical
large-scale structure covariances, and the luminosity and stellar mass function
of galaxies in clusters, as well as the radial profile of the projected
distribution of galaxies.  
Clustering statistics of clusters are also provided, including the
cluster-cluster \gls{2-pcf} and its associated
covariance matrix.

\section{IAL detailed design and features}
\label{app:ial}
This appendix provides additional details on the internal design and features of
the \gls{ial}, focusing on the Meta-Scheduler and the Pipeline-Runner.

\subsection{Meta-Scheduler}

\begin{figure*}[!t]
    \centering
    \includegraphics[width=1\linewidth]{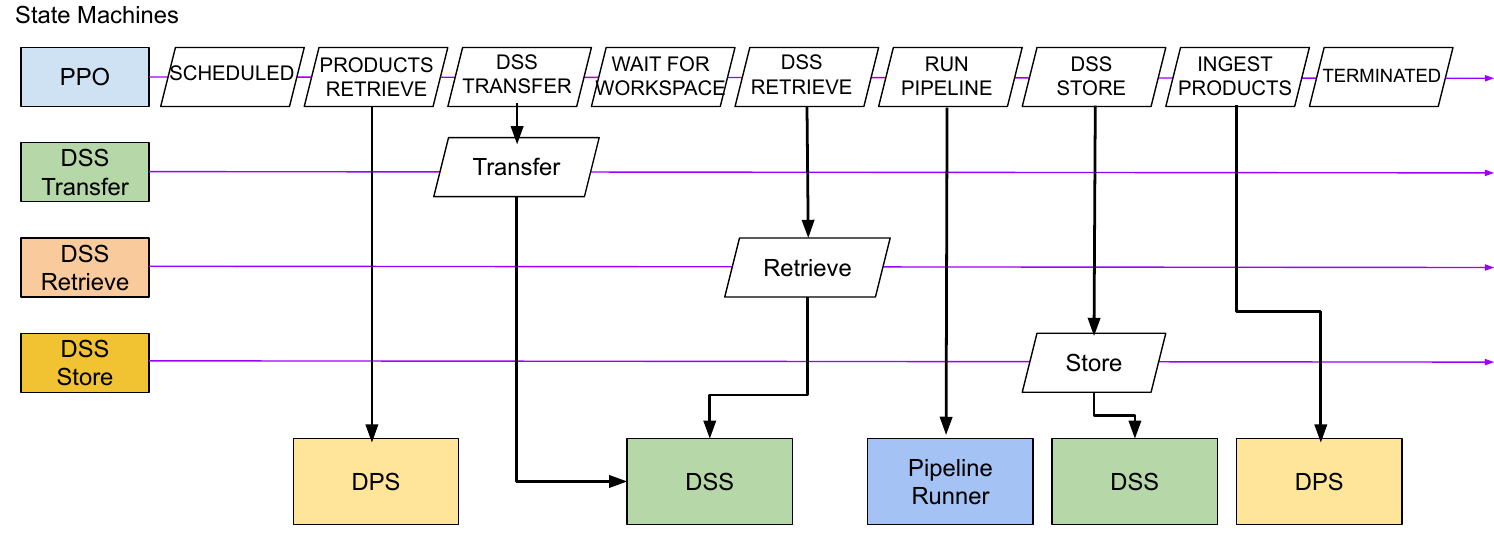}
    \caption{High-level view of states managed by different state machines inside Meta-Scheduler.}
    \label{SGS_MS_Overview}
\end{figure*}

\subsubsection{Architecture and responsibilities}

The main responsibility of the Meta-Scheduler is to schedule \glspl{ppo} on a
given \gls{sdc} in the most optimal way to ensure efficient processing of the
different \glspl{ppo}. It manages communications and queueing with external
services like \gls{dss} and \gls{dps}. For this reason, each \gls{ppo} has to
define its maximal estimated workspace usage during processing. 

Internally, Meta-Scheduler is made up of multiple state machines managing the
flow of \glspl{ppo}. These state machines have a 1:1 mapping between execution
state and a Task (java implementation of what do to in this state).
Meta-Scheduler manages the capacity of the different states in those state
machines and ensures optimal resource usage through prioritisation, resource
management and caching of files. For example the number of \glspl{ppo} that can
be retrieving products at the same time from \gls{dps} is limited.

There are four main state machines: \texttt{\gls{ppo}},
\texttt{\gls{dss}-Transfer}, \texttt{\gls{dss}-Retrieve}, and
\texttt{\gls{dss}-Store}. All the elements inside a state machine use a
prioritisation logic to minimise the effect of bottlenecks and to ensure
efficient processing of the different \glspl{ppo}. See
Fig.~\ref{SGS_MS_Overview} for a high-level view of the different states and
state-machines from Meta-Schedulerg.

\subsubsection{PPO processing and caching}

\begin{figure}[!htbp]
    \centering
    \includegraphics[width=1\linewidth]{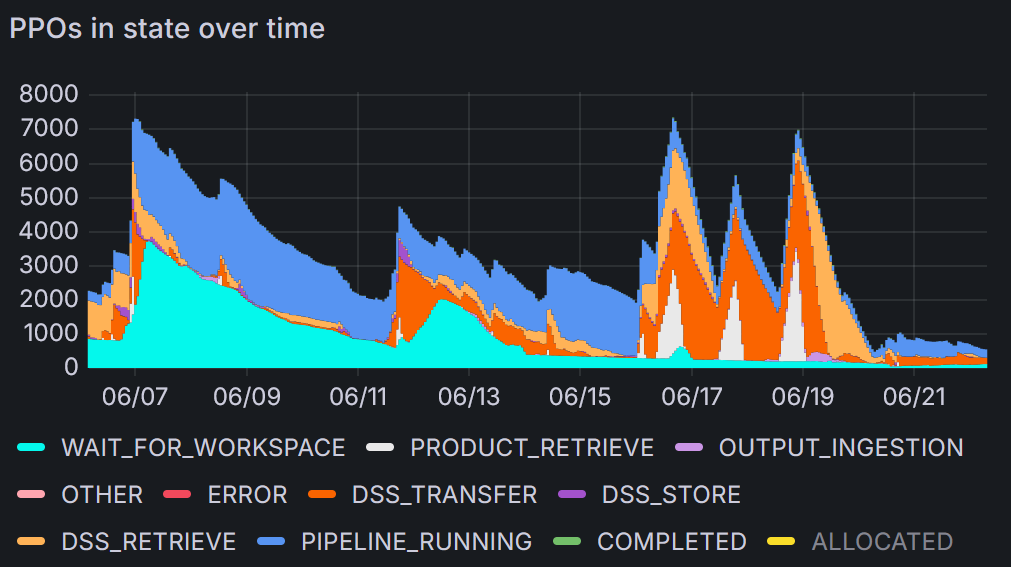}
    \caption{\gls{ppo} processing states of the \gls{sgs} in June 2025 as seen in the global Grafana dashboard.}
    \label{SGS_processing}
\end{figure}

The \texttt{\gls{ppo}} state machine manages the states of the individual
\glspl{ppo}. Every \gls{ppo} has a list of input files it requires for
processing. This list of files is split into smaller chunks, which are then
first requested to be made locally available on the local \gls{dss} instance of
the \gls{sdc} through the \texttt{\gls{dss}-Transfer} state machine. Once all
files are locally available on \gls{dss}, Meta-Scheduler will request \gls{dss}
to transfer these files to the local workspace cache of Meta-Scheduler using the
\texttt{\gls{dss}-Retrieve} state machine. Once all files are copied to the
Workspace cache, execution of the processing pipeline is triggered with the
\texttt{Pipeline-Runner}. After processing completion, Meta-Scheduler ingests
the output files in chunks using the \texttt{\gls{dss}-Store} state machine.
Finally, Meta-Scheduler ingests the output products into \gls{dps} and
terminates the \gls{ppo}. The progression of a \gls{ppo} is reflected in states,
as shown in Figs. \ref{SGS_MS_Overview} and \ref{SGS_processing}.

One important bottleneck in the processing pipelines of \Euclid is the large
number of files. Having a local cache of files requested by \glspl{ppo} speeds
up the processing and the ramp up time of consecutive \glspl{ppo} requesting the
same files. Managing this cache is a core responsibility of the Meta-Scheduler.

\subsection{Pipeline-Runner}

\begin{figure*}[!t]
\centering
\begin{minipage}{0.98\textwidth}
\begin{lstlisting}[style=pythonpaper]
from euclidwf.framework.workflow_dsl import pipeline, parallel
from euclidwf.framework.taskdefs import Executable, Input, Output, ComputingResources

cutout = Executable(command="E-Run ST_SHE 1.0.1 galaxy_cutout --mode=HDF5",
   inputs=[Input("catalogue"), Input("data_files", content_type="listfile")],
   outputs=[Output("cutout_list", content_type="listfile")],
   resources=ComputingResources(cores=1, ram=2.0, walltime=0.5) )

shear_processing = Executable( command="E-Run ST_SHE 1.0.1 shear_processing", 
   inputs=[Input("cutout_image"), Input("config")],
   outputs=[Output("shear")], 
   resources=ComputingResources(cores=1, ram=1.0, walltime=1.0) )

@parallel(iterable='cutout')
def parallel_shear(cutout, config_product):
   shear = shear_processing(cutout_image=cutout, config=config_product)
   return shear

@pipeline
def example_pipeline(catalogue, data_files, config_product):
   cutout_list = cutout(catalogue=catalogue, data_files=data_files)
   shears = parallel_shear(cutout=cutout_list, config_product=config_product)
   return shears
\end{lstlisting}
\end{minipage}
\caption{Example Pipeline Framework definition.}
\label{lst:pipeline-framework}
\end{figure*}

\subsubsection{Architecture and pilot model}

The Pipeline-Runner is implemented as a distributed microservice architecture
using \texttt{ZeroMQ} for communication. It employs both publish--subscribe and
request--reply patterns. The former is used for broadcasting messages (e.g., to
pilots), while the latter supports direct communication between individual
services.

Pilot job specifications (\gls{cpu} cores, virtual memory, local disk,
\glspl{gpu}, resident memory, and wall time) are defined by system
administrators to match the characteristics of the underlying clusters. The
Pipeline-Runner supports multiple Distributed Resource Manager systems,
including Slurm, Portable Batch System, Sun Grid Engine, HTCondor, and local
process execution. Two main deployment modes are in use. On dedicated clusters,
pilots typically occupy full or half nodes. On shared clusters, pilot sizes vary
from small (e.g., 1 core, 4\,GB of resident memory) to large configurations. The
Pipeline-Runner optimises task assignment to pilots while monitoring the
Distributed Resource Manager queue, favouring pilots that can start quickly on
shared resources.

Pilots execute multiple \gls{pf} tasks and perform continuous profiling of each
task and of the pilot itself. Metrics such as CPU usage, resident memory,
virtual memory, and \gls{i/o} activity are collected at regular intervals
(typically every 15 seconds). Profiling data are aggregated at multiple levels:
individual task; task type; pipeline; and cluster. These data are exposed
through real-time visualisations in the Pipeline-Runner web interface and in the
\gls{sgs}-wide monitoring dashboards. After completion of a \gls{ppo}, the
collected profiling data are analysed and summarised in a report. This report
provides insights into resource usage, including comparisons between requested
and actual resource consumption, and helps identify inefficiencies and outliers.
For output files that are later ingested into the \gls{dss}, pilots compute
checksums to improve cache utilisation on the shared file system.

The scheduling system operates at large scale, handling millions of pending
tasks across thousands of \glspl{ppo} and up to tens of thousands of pilots.
Task-to-pilot assignment is based on a scoring algorithm that considers multiple
factors, including fairness across pipelines, resource requirements, \gls{i/o}
pressure, local disk usage, and task age. System administrators can tune the
scheduling behaviour by adjusting weights assigned to these factors. This
allows, for example, preferential execution of memory-intensive tasks on
high-memory nodes.

\subsubsection{Framework}

The Pipeline Framework provides a Python-based domain-specific language to
define pipelines as data-flow graphs. Pipeline steps are defined as executable
units with explicit inputs, outputs, and resource requirements. Pipelines are
composed using decorators that define execution semantics such as sequential and
parallel processing.

Figure \ref{lst:pipeline-framework} shows a simple example pipeline. This
example shows how developers define tasks, specify resource requirements, and
express parallelism over collections of data. The framework translates this
definition into a data-flow graph that is evaluated dynamically by the
Pipeline-Runner. Lines 4--7 and 9--12 define executable units. Each executable
must comply with the \gls{ial} Task application programming interface. All
inputs and outputs are passed via command-line arguments (e.g.
\texttt{----catalogue}, \texttt{----data\_files}, \texttt{----cutout\_list}),
with paths provided by the Pipeline-Runner. The Task application programming
interface further defines standard arguments such as \texttt{----workdir}, which
points to the \gls{ppo} working directory on the shared file system, and
\texttt{----logdir}, which specifies the subdirectory where the task logs are
written. Each executable specifies its resource requirements (CPU, memory, disk
space, GPUs, and wall time). Resource specifications can also be defined
dynamically during execution, for example by deriving memory requirements from
the output of a preceding task.

The main entry point of the pipeline is defined by the \texttt{@pipeline}
decorator. Input data are prepared by the Meta-Scheduler in production, or
provided by the user in development environments. The function defined with the
\texttt{@parallel} decorator implements a map-reduce style parallelisation. The
number of parallel branches is determined at runtime based on the input data
(e.g., the number of cutouts produced by the first task). The outputs of the
parallel branches are merged into a single file containing references to the
individual results. The parallel construct supports multiple iterators,
different ordering and merging strategies, and nested execution.

\subsubsection{Lineage tracking, scalability, and robustness}

The Pipeline-Runner tracks data lineage from input files to output products.
This information is ingested into the \gls{dps} lineage service and enables
tracing dependencies across \gls{ppo} boundaries at the \gls{sgs} level.

The messaging system supports back-pressure mechanisms, adapting communication
rates under high load conditions. All I/O operations are designed to tolerate
extreme latency in the shared file system. Under heavy load, file operations
such as writes or directory listings may be blocked for extended periods. The
\gls{ial} handles these conditions explicitly to avoid deadlocks and to ensure
continued orchestration during large processing campaigns. Several retry
strategies are implemented to improve robustness in a large-scale distributed
environment. Transient failures, in particular those caused by infrastructure
issues, are detected and handled automatically to avoid unnecessary \gls{ppo}
failures.

\section{SGS technical budget}
\label{app:SGS technical budget}
\begin{figure*}
    \centering
    \includegraphics[width=1\linewidth]{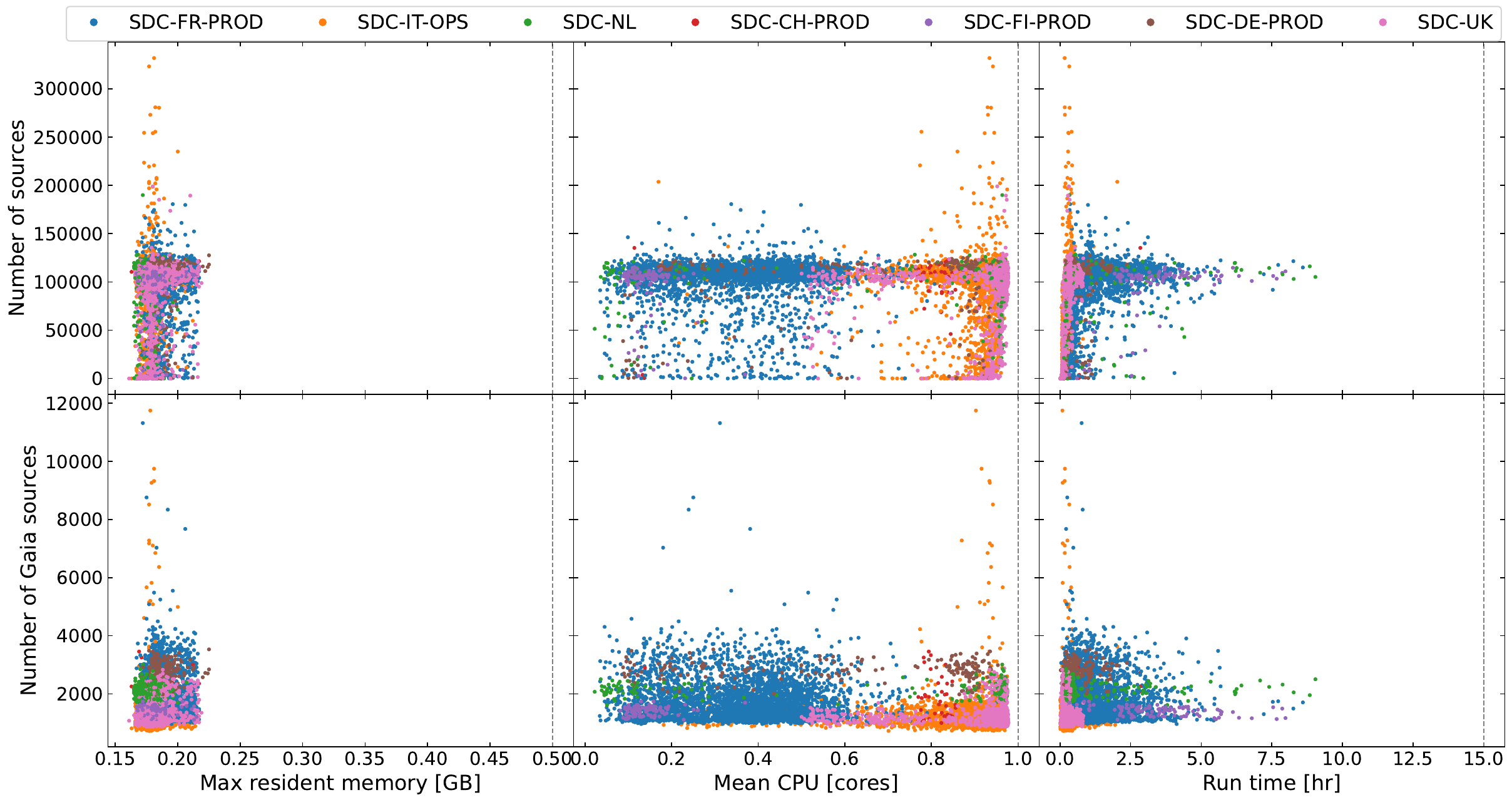}
    \caption{Resource usage of the MER initialise task during the \gls{dr1}
    \gls{ews} processing. Each point corresponds to one execution of the task on
    a single tile, coloured by the \gls{sdc} on which it was run. The columns
    show the resource usage: the peak memory (max resident memory), the mean number of
    used CPU cores, and the total run time (from left to right), as a function
    of the number of sources in the tile (top row) and the number of Gaia
    sources (bottom row).}
    \label{MER_profiling}
\end{figure*}

As many \glspl{pf}, each composed of multiple tasks, run within the \gls{sgs},
it is essential to profile them individually to assess performance and optimise
code when necessary, ensuring efficient use of resources available in the
\glspl{sdc}. The performance of each \gls{pf} depends both on the tasks it
executes and on its specific objectives. To estimate the \gls{sgs} technical
budget, performance metrics such as CPU and RAM usage are computed and recorded
for each task, for each \gls{pf}. This profiling is carried out thanks to the
work of the \gls{ial}, and monitoring and control teams. An example of profiled
elements is shown in Fig.~\ref{MER_profiling}, for resources (CPU, resident
memory, and run time) used during the \gls{dr1} \gls{ews} processing for the MER
initialise task.

The MER \gls{pf}, used here as an illustration, consists of tens of tasks, each
with distinct requirements in terms of memory, CPU, and I/O. Only one task is
illustrated here. Resource needs may also vary depending on the sources within
the tile to be processed. In this context, \gls{pf} developers have worked to
optimise resource allocation within the \glspl{sdc} by identifying tasks with
overestimated requirements and refining requests to better match actual needs.
Profiling across tasks and \glspl{pf} has enabled the computation of a reliable
processing budget, demonstrating our ability to accurately evaluate compute
needs.

\newpage
\section{Version of PFs used for DR1}
\label{app:version PFs}
\begin{table}[htp!]
    \centering
    \caption{Project versions of the PFs used for DR1.}
    \label{tab:processing_versions}
    \begin{tabular}{ll}
        \hline\hline
        Processing Function & Project Version \\
        \hline
        EXT         & 0.12.9  \\
        VIS         & 18.0.4  \\
        NIR         & 6.1.2   \\
        MER         & 11.3.1  \\
        SIR         & 6.0.2   \\
        LE3-ID-VMPZ & 1.6.4 and 1.6.5   \\
        SPE         & 25.0.1  \\
        PHZ         & 2.16.13 \\
        SHE         & 9.12.2  \\
        \hline
    \end{tabular}
\end{table}

\end{appendix}

\label{LastPage}
\end{document}

%% file: acronyms.tex
\newacronym{2-pcf}{2-PCF}{2-point correlation function}
\newacronym{3-pcf}{3-PCF}{3-point correlation function}
\newacronym{ares}{ARES}{Analysis and Reporting System}
\newacronym{cdm}{CDM}{Common Data Model}
\newacronym{ch}{CH}{Switzerland}
\newacronym{cl}{CL}{clusters of galaxies}
\newacronym{codeen}{CODEEN}{Collaborative Development Environment}
\newacronym{coors}{COORS}{COmmon ORchestration System}
\newacronym{cpu}{CPU}{Central Processing Unit}
\newacronym{cvmfs}{CVMFS}{CernVM File System}
\newacronym{de}{DE}{Germany}
\newacronym{dpdd}{DPDD}{Data Product Description Document}
\newacronym{dps}{DPS}{Data Processing System}
\newacronym{dr1}{DR1}{Data Release 1}
\newacronym{dss}{DSS}{Distributed Storage System}
\newacronym{eas}{EAS}{Euclid Archive System}
\newacronym{ec}{EC}{Euclid Consortium}
\newacronym{ecb}{ECB}{Euclid Consortium Board}
\newacronym{eden}{EDEN}{Euclid Development Environment}
\newacronym{edf}{EDF}{Euclid Deep Field}
\newacronym{edff}{EDF-F}{Euclid Deep Field Fornax}
\newacronym{edfn}{EDF-N}{Euclid Deep Field North}
\newacronym{edfs}{EDF-S}{Euclid Deep Field South}
\newacronym{eds}{EDS}{Euclid Deep Survey}
\newacronym{es}{ES}{Spain}
\newacronym{esac}{ESAC}{European Space Astronomy Centre}
\newacronym{ews}{EWS}{Euclid Wide Survey}
\newacronym{ext}{EXT}{External}
\newacronym{fi}{FI}{Finland}
\newacronym{fits}{FITS}{Flexible Image Transport System}
\newacronym{fr}{FR}{France}
\newacronym{gc}{GC}{galaxy clustering}
\newacronym{gpu}{GPU}{Graphics Processing Unit}
\newacronym{gui}{GUI}{Graphical User Interface}
\newacronym{hktm}{HKTM}{housekeeping telemetry}
\newacronym{htc}{HTC}{High Throughput Computing}
\newacronym{https}{HTTPS}{Hypertext Transfer Protocol Secure}
\newacronym{ial}{IAL}{Infrastructure Abstraction Layer}
\newacronym{i/o}{I/O}{Input/Output}
\newacronym{ioda}{IODA}{Instrument Operations Data Analysis}
\newacronym{iot}{IOT}{Instrument Operations Team}
\newacronym{it}{IT}{Italy}
\newacronym{le1}{LE1}{Level 1}
\newacronym{le2}{LE2}{Level 2}
\newacronym{le3}{LE3}{Level 3}
\newacronym{lodeen}{LODEEN}{Local Development Environment}
\newacronym{mer}{MER}{Merge}
\newacronym{moc}{MOC}{Mission Operations Centre}
\newacronym{mpi}{MPI}{Message Passing Interface}
\newacronym{nir}{NIR}{Near-infrared}
\newacronym{nisp}{NISP}{Near-Infrared Spectrophotometer}
\newacronym{nl}{NL}{Netherlands}
\newacronym{oss}{OSS}{Operation Sky Survey}
\newacronym{otf}{OTF}{on-the-fly}
\newacronym{ou}{OU}{Organisation Unit}
\newacronym{pf}{PF}{Processing Function}
\newacronym{phz}{PHZ}{Photometric redshift}
\newacronym{ppo}{PPO}{Pipeline Processing Order}
\newacronym{qla}{QLA}{Quick Look Analysis}
\newacronym{rest api}{REST API}{Representational State Transfer Application Programming Interface}
\newacronym{ros}{ROS}{Reference Observation Sequence}
\newacronym{sdc}{SDC}{Science Data Centre}
\newacronym{sgs}{SGS}{Science Ground Segment}
\newacronym{she}{SHE}{Shear}
\newacronym{sim}{SIM}{Simulations}
\newacronym{sir}{SIR}{Spectroscopic reduction}
\newacronym{soc}{SOC}{Science Operations Centre}
\newacronym{spe}{SPE}{Spectroscopic redshift}
\newacronym{unions}{UNIONS}{Ultraviolet Near-Infrared Optical Northern Survey}
\newacronym{uk}{UK}{United Kingdom}
\newacronym{us}{US}{United States}
\newacronym{vis}{VIS}{Visible}
\newacronym{wl}{WL}{weak gravitational lensing}
\newacronym{xml}{XML}{Extensible Markup Language}

%% file: authors.tex
\newcommand{\orcid}[1]{} 
\author{Euclid Collaboration: M.~Ruffenach\orcid{0000-0001-8772-7330}\inst{\ref{aff1}}
\and M.~Frailis\orcid{0000-0002-7400-2135}\thanks{\email{marco.frailis@inaf.it}}\inst{\ref{aff2}}
\and A.~Zacchei\orcid{0000-0003-0396-1192}\inst{\ref{aff2},\ref{aff3}}
\and P.~Casenove\orcid{0009-0006-6736-1670}\inst{\ref{aff1}}
\and M.~S.~Holliman\orcid{0009-0006-9850-9770}\inst{\ref{aff4}}
\and H.~Aussel\orcid{0000-0002-1371-5705}\inst{\ref{aff5}}
\and B.~Altieri\orcid{0000-0003-3936-0284}\inst{\ref{aff6}}
\and G.~Angora\orcid{0000-0002-0316-6562}\inst{\ref{aff7},\ref{aff8}}
\and A.~Basset\orcid{0009-0001-5490-3139}\inst{\ref{aff1}}
\and A.~N.~Belikov\inst{\ref{aff9},\ref{aff10}}
\and J.~Benarroch\inst{\ref{aff11}}
\and T.~Bouvard\orcid{0009-0002-7959-312X}\inst{\ref{aff11}}
\and M.~Brescia\orcid{0000-0001-9506-5680}\inst{\ref{aff12},\ref{aff7}}
\and S.~Cavuoti\orcid{0000-0002-3787-4196}\inst{\ref{aff7},\ref{aff13}}
\and S.~Farrens\orcid{0000-0002-9594-9387}\inst{\ref{aff5}}
\and R.~Gimenez\orcid{0009-0005-7181-7071}\inst{\ref{aff1}}
\and R.~Giusteri\orcid{0000-0001-9793-5037}\inst{\ref{aff2}}
\and X.~Goudot\inst{\ref{aff11}}
\and A.~Gregorio\orcid{0000-0003-4028-8785}\inst{\ref{aff14},\ref{aff2},\ref{aff15}}
\and J.~Haase\orcid{0000-0001-7809-3671}\inst{\ref{aff16}}
\and T.~Heuberger\orcid{0009-0009-3474-8467}\inst{\ref{aff17}}
\and S.~Kruk\orcid{0000-0001-8010-8879}\inst{\ref{aff6}}
\and G.~Libet\inst{\ref{aff1}}
\and S.~Marcin\inst{\ref{aff17}}
\and S.~Maurogordato\inst{\ref{aff18}}
\and M.~Melchior\inst{\ref{aff19}}
\and F.~Pasian\orcid{0000-0002-4869-3227}\inst{\ref{aff2}}
\and G.~Riccio\inst{\ref{aff7}}
\and F.~Rizzo\orcid{0000-0002-9407-585X}\inst{\ref{aff2}}
\and E.~A.~Valentijn\inst{\ref{aff9}}
\and O.~R.~Williams\orcid{0000-0003-0274-1526}\inst{\ref{aff20}}
\and S.~Andreon\orcid{0000-0002-2041-8784}\inst{\ref{aff21}}
\and N.~Auricchio\orcid{0000-0003-4444-8651}\inst{\ref{aff22}}
\and S.~Awan\inst{\ref{aff23}}
\and C.~Baccigalupi\orcid{0000-0002-8211-1630}\inst{\ref{aff3},\ref{aff2},\ref{aff15},\ref{aff24}}
\and M.~Baldi\orcid{0000-0003-4145-1943}\inst{\ref{aff25},\ref{aff22},\ref{aff26}}
\and A.~Balestra\orcid{0000-0002-6967-261X}\inst{\ref{aff27}}
\and S.~Bardelli\orcid{0000-0002-8900-0298}\inst{\ref{aff22}}
\and P.~Battaglia\orcid{0000-0002-7337-5909}\inst{\ref{aff22}}
\and R.~Bender\orcid{0000-0001-7179-0626}\inst{\ref{aff16},\ref{aff28}}
\and A.~Biviano\orcid{0000-0002-0857-0732}\inst{\ref{aff2},\ref{aff3}}
\and R.~P.~Blake\inst{\ref{aff4}}
\and A.~Bonchi\orcid{0000-0002-2667-5482}\inst{\ref{aff29}}
\and E.~Branchini\orcid{0000-0002-0808-6908}\inst{\ref{aff30},\ref{aff31},\ref{aff21}}
\and S.~Camera\orcid{0000-0003-3399-3574}\inst{\ref{aff32},\ref{aff33},\ref{aff34}}
\and V.~Capobianco\orcid{0000-0002-3309-7692}\inst{\ref{aff34}}
\and C.~Carbone\orcid{0000-0003-0125-3563}\inst{\ref{aff35}}
\and J.~Carretero\orcid{0000-0002-3130-0204}\inst{\ref{aff36},\ref{aff37}}
\and M.~Castellano\orcid{0000-0001-9875-8263}\inst{\ref{aff38}}
\and G.~Castignani\orcid{0000-0001-6831-0687}\inst{\ref{aff22}}
\and P.-Y.~Chabaud\inst{\ref{aff39}}
\and K.~C.~Chambers\orcid{0000-0001-6965-7789}\inst{\ref{aff40}}
\and A.~Cimatti\inst{\ref{aff41}}
\and C.~Colodro-Conde\inst{\ref{aff42}}
\and G.~Congedo\orcid{0000-0003-2508-0046}\inst{\ref{aff4}}
\and C.~J.~Conselice\orcid{0000-0003-1949-7638}\inst{\ref{aff43}}
\and L.~Conversi\orcid{0000-0002-6710-8476}\inst{\ref{aff44},\ref{aff6}}
\and Y.~Copin\orcid{0000-0002-5317-7518}\inst{\ref{aff45}}
\and A.~Costille\inst{\ref{aff39}}
\and F.~Courbin\orcid{0000-0003-0758-6510}\inst{\ref{aff46},\ref{aff47},\ref{aff48}}
\and H.~M.~Courtois\orcid{0000-0003-0509-1776}\inst{\ref{aff49}}
\and M.~Cropper\orcid{0000-0003-4571-9468}\inst{\ref{aff23}}
\and R.~da~Silva\orcid{0000-0003-4788-677X}\inst{\ref{aff38},\ref{aff29}}
\and H.~Degaudenzi\orcid{0000-0002-5887-6799}\inst{\ref{aff50}}
\and S.~de~la~Torre\inst{\ref{aff39}}
\and G.~De~Lucia\orcid{0000-0002-6220-9104}\inst{\ref{aff2}}
\and C.~Dolding\orcid{0009-0003-7199-6108}\inst{\ref{aff23}}
\and H.~Dole\orcid{0000-0002-9767-3839}\inst{\ref{aff51}}
\and F.~Dubath\orcid{0000-0002-6533-2810}\inst{\ref{aff50}}
\and X.~Dupac\inst{\ref{aff6}}
\and S.~Dusini\orcid{0000-0002-1128-0664}\inst{\ref{aff52}}
\and S.~Escoffier\orcid{0000-0002-2847-7498}\inst{\ref{aff53}}
\and M.~Farina\orcid{0000-0002-3089-7846}\inst{\ref{aff54}}
\and R.~Farinelli\inst{\ref{aff22}}
\and F.~Faustini\orcid{0000-0001-6274-5145}\inst{\ref{aff38}}
\and S.~Ferriol\inst{\ref{aff45}}
\and F.~Finelli\orcid{0000-0002-6694-3269}\inst{\ref{aff22},\ref{aff55}}
\and N.~Fourmanoit\orcid{0009-0005-6816-6925}\inst{\ref{aff53}}
\and E.~Franceschi\orcid{0000-0002-0585-6591}\inst{\ref{aff22}}
\and P.~Franzetti\inst{\ref{aff35}}
\and M.~Fumana\orcid{0000-0001-6787-5950}\inst{\ref{aff35}}
\and L.~Gabarra\orcid{0000-0002-8486-8856}\inst{\ref{aff56}}
\and S.~Galeotta\orcid{0000-0002-3748-5115}\inst{\ref{aff2}}
\and K.~George\orcid{0000-0002-1734-8455}\inst{\ref{aff57}}
\and W.~Gillard\orcid{0000-0003-4744-9748}\inst{\ref{aff53}}
\and B.~Gillis\orcid{0000-0002-4478-1270}\inst{\ref{aff4}}
\and C.~Giocoli\orcid{0000-0002-9590-7961}\inst{\ref{aff22},\ref{aff26}}
\and P.~G\'omez-Alvarez\orcid{0000-0002-8594-5358}\inst{\ref{aff58},\ref{aff6}}
\and J.~Gracia-Carpio\orcid{0000-0003-4689-3134}\inst{\ref{aff16}}
\and A.~Grazian\orcid{0000-0002-5688-0663}\inst{\ref{aff27}}
\and F.~Grupp\inst{\ref{aff16},\ref{aff28}}
\and S.~Gwyn\orcid{0000-0001-8221-8406}\inst{\ref{aff59}}
\and W.~G.~Hartley\orcid{0000-0002-4563-5101}\inst{\ref{aff50}}
\and S.~V.~H.~Haugan\orcid{0000-0001-9648-7260}\inst{\ref{aff60}}
\and G.~Helou\orcid{0000-0003-3367-3415}\inst{\ref{aff61},\ref{aff62}}
\and S.~Hemmati\orcid{0000-0003-2226-5395}\inst{\ref{aff62}}
\and J.~Hoar\inst{\ref{aff6}}
\and H.~Hoekstra\orcid{0000-0002-0641-3231}\inst{\ref{aff63}}
\and W.~Holmes\orcid{0009-0007-8554-4646}\inst{\ref{aff64}}
\and I.~M.~Hook\orcid{0000-0002-2960-978X}\inst{\ref{aff65}}
\and F.~Hormuth\inst{\ref{aff66}}
\and A.~Hornstrup\orcid{0000-0002-3363-0936}\inst{\ref{aff67},\ref{aff68}}
\and K.~Jahnke\orcid{0000-0003-3804-2137}\inst{\ref{aff69}}
\and M.~Jhabvala\inst{\ref{aff70}}
\and B.~Joachimi\orcid{0000-0001-7494-1303}\inst{\ref{aff71}}
\and S.~Kermiche\orcid{0000-0002-0302-5735}\inst{\ref{aff53}}
\and A.~Kiessling\orcid{0000-0002-2590-1273}\inst{\ref{aff64}}
\and R.~Kohley\inst{\ref{aff6}}
\and B.~Kubik\orcid{0009-0006-5823-4880}\inst{\ref{aff45}}
\and M.~K\"ummel\orcid{0000-0003-2791-2117}\inst{\ref{aff28}}
\and M.~Kunz\orcid{0000-0002-3052-7394}\inst{\ref{aff72}}
\and H.~Kurki-Suonio\orcid{0000-0002-4618-3063}\inst{\ref{aff73},\ref{aff74}}
\and R.~Laureijs\inst{\ref{aff9}}
\and Q.~Le~Boulc'h\inst{\ref{aff75}}
\and A.~M.~C.~Le~Brun\orcid{0000-0002-0936-4594}\inst{\ref{aff76}}
\and V.~Le~Brun\orcid{0000-0002-5027-1939}\inst{\ref{aff39}}
\and D.~Le~Mignant\orcid{0000-0002-5339-5515}\inst{\ref{aff39}}
\and P.~Liebing\orcid{0009-0005-8043-7150}\inst{\ref{aff23}}
\and S.~Ligori\orcid{0000-0003-4172-4606}\inst{\ref{aff34}}
\and P.~B.~Lilje\orcid{0000-0003-4324-7794}\inst{\ref{aff60}}
\and V.~Lindholm\orcid{0000-0003-2317-5471}\inst{\ref{aff73},\ref{aff74}}
\and I.~Lloro\orcid{0000-0001-5966-1434}\inst{\ref{aff77}}
\and M.~Magliocchetti\orcid{0000-0001-9158-4838}\inst{\ref{aff54}}
\and G.~Mainetti\orcid{0000-0003-2384-2377}\inst{\ref{aff75}}
\and D.~Maino\inst{\ref{aff78},\ref{aff35},\ref{aff79}}
\and O.~Mansutti\orcid{0000-0001-5758-4658}\inst{\ref{aff2}}
\and O.~Marggraf\orcid{0000-0001-7242-3852}\inst{\ref{aff80}}
\and M.~Martinelli\orcid{0000-0002-6943-7732}\inst{\ref{aff38},\ref{aff81}}
\and N.~Martinet\orcid{0000-0003-2786-7790}\inst{\ref{aff39}}
\and F.~Marulli\orcid{0000-0002-8850-0303}\inst{\ref{aff82},\ref{aff22},\ref{aff26}}
\and R.~J.~Massey\orcid{0000-0002-6085-3780}\inst{\ref{aff83}}
\and N.~Mauri\orcid{0000-0001-8196-1548}\inst{\ref{aff41},\ref{aff26}}
\and H.~J.~McCracken\orcid{0000-0002-9489-7765}\inst{\ref{aff84}}
\and C.~J.~R.~McPartland\orcid{0000-0003-0639-025X}\inst{\ref{aff68},\ref{aff85}}
\and E.~Medinaceli\orcid{0000-0002-4040-7783}\inst{\ref{aff22}}
\and S.~Mei\orcid{0000-0002-2849-559X}\inst{\ref{aff86},\ref{aff87}}
\and M.~Meneghetti\orcid{0000-0003-1225-7084}\inst{\ref{aff22},\ref{aff26}}
\and E.~Merlin\orcid{0000-0001-6870-8900}\inst{\ref{aff27}}
\and G.~Meylan\orcid{0000-0001-6503-0209}\inst{\ref{aff88}}
\and J.~J.~Mohr\orcid{0000-0002-6875-2087}\inst{\ref{aff57}}
\and P.~Monaco\orcid{0000-0003-2083-7564}\inst{\ref{aff14},\ref{aff2},\ref{aff15},\ref{aff3}}
\and A.~Mora\orcid{0000-0002-1922-8529}\inst{\ref{aff89}}
\and M.~Moresco\orcid{0000-0002-7616-7136}\inst{\ref{aff82},\ref{aff22}}
\and C.~Moretti\orcid{0000-0003-3314-8936}\inst{\ref{aff2},\ref{aff3},\ref{aff15}}
\and B.~Morin\inst{\ref{aff5},\ref{aff90}}
\and P.~W.~Morris\orcid{0000-0002-5186-4381}\inst{\ref{aff61}}
\and L.~Moscardini\orcid{0000-0002-3473-6716}\inst{\ref{aff82},\ref{aff22},\ref{aff26}}
\and E.~Munari\orcid{0000-0002-1751-5946}\inst{\ref{aff2},\ref{aff3}}
\and R.~Nakajima\orcid{0009-0009-1213-7040}\inst{\ref{aff80}}
\and C.~Neissner\orcid{0000-0001-8524-4968}\inst{\ref{aff91},\ref{aff37}}
\and R.~C.~Nichol\orcid{0000-0003-0939-6518}\inst{\ref{aff92}}
\and S.-M.~Niemi\orcid{0009-0005-0247-0086}\inst{\ref{aff93}}
\and C.~Padilla\orcid{0000-0001-7951-0166}\inst{\ref{aff91}}
\and S.~Paltani\orcid{0000-0002-8108-9179}\inst{\ref{aff50}}
\and W.~J.~Percival\orcid{0000-0002-0644-5727}\inst{\ref{aff94},\ref{aff95},\ref{aff96}}
\and V.~Pettorino\orcid{0000-0002-4203-9320}\inst{\ref{aff93}}
\and A.~Pezzotta\orcid{0000-0003-0726-2268}\inst{\ref{aff21}}
\and S.~Pires\orcid{0000-0002-0249-2104}\inst{\ref{aff5}}
\and G.~Polenta\orcid{0000-0003-4067-9196}\inst{\ref{aff29}}
\and M.~Poncet\inst{\ref{aff1}}
\and L.~A.~Popa\inst{\ref{aff97}}
\and L.~Pozzetti\orcid{0000-0001-7085-0412}\inst{\ref{aff22}}
\and G.~D.~Racca\orcid{0000-0002-9883-8981}\inst{\ref{aff63},\ref{aff93}}
\and F.~Raison\orcid{0000-0002-7819-6918}\inst{\ref{aff16}}
\and R.~Rebolo\orcid{0000-0003-3767-7085}\inst{\ref{aff42},\ref{aff98},\ref{aff99}}
\and A.~Renzi\orcid{0000-0001-9856-1970}\inst{\ref{aff100},\ref{aff52},\ref{aff22}}
\and J.~Rhodes\orcid{0000-0002-4485-8549}\inst{\ref{aff64}}
\and I.~Risso\orcid{0000-0003-2525-7761}\inst{\ref{aff30},\ref{aff31},\ref{aff21}}
\and E.~Romelli\orcid{0000-0003-3069-9222}\inst{\ref{aff2}}
\and M.~Roncarelli\orcid{0000-0001-9587-7822}\inst{\ref{aff22}}
\and C.~Rosset\orcid{0000-0003-0286-2192}\inst{\ref{aff86}}
\and B.~Rusholme\orcid{0000-0001-7648-4142}\inst{\ref{aff62}}
\and R.~Saglia\orcid{0000-0003-0378-7032}\inst{\ref{aff28},\ref{aff16}}
\and Z.~Sakr\orcid{0000-0002-4823-3757}\inst{\ref{aff101},\ref{aff102},\ref{aff103}}
\and A.~G.~S\'anchez\orcid{0000-0003-1198-831X}\inst{\ref{aff16}}
\and D.~Sapone\orcid{0000-0001-7089-4503}\inst{\ref{aff104}}
\and B.~Sartoris\orcid{0000-0003-1337-5269}\inst{\ref{aff16},\ref{aff2}}
\and M.~Sauvage\orcid{0000-0002-0809-2574}\inst{\ref{aff5}}
\and P.~Schneider\orcid{0000-0001-8561-2679}\inst{\ref{aff80}}
\and T.~Schrabback\orcid{0000-0002-6987-7834}\inst{\ref{aff105}}
\and M.~Scodeggio\inst{\ref{aff35}}
\and A.~Secroun\orcid{0000-0003-0505-3710}\inst{\ref{aff53}}
\and E.~Sihvola\orcid{0000-0003-1804-7715}\inst{\ref{aff106}}
\and C.~Sirignano\orcid{0000-0002-0995-7146}\inst{\ref{aff100},\ref{aff52}}
\and G.~Sirri\orcid{0000-0003-2626-2853}\inst{\ref{aff26}}
\and L.~Stanco\orcid{0000-0002-9706-5104}\inst{\ref{aff52}}
\and C.~Surace\orcid{0000-0003-2592-0113}\inst{\ref{aff39}}
\and P.~Tallada-Cresp\'{i}\orcid{0000-0002-1336-8328}\inst{\ref{aff36},\ref{aff37}}
\and A.~N.~Taylor\inst{\ref{aff4}}
\and H.~I.~Teplitz\orcid{0000-0002-7064-5424}\inst{\ref{aff107}}
\and I.~Tereno\orcid{0000-0002-4537-6218}\inst{\ref{aff108},\ref{aff109}}
\and N.~Tessore\orcid{0000-0002-9696-7931}\inst{\ref{aff23}}
\and S.~Toft\orcid{0000-0003-3631-7176}\inst{\ref{aff110},\ref{aff85}}
\and R.~Toledo-Moreo\orcid{0000-0002-2997-4859}\inst{\ref{aff111},\ref{aff112}}
\and F.~Torradeflot\orcid{0000-0003-1160-1517}\inst{\ref{aff37},\ref{aff36}}
\and A.~Tsyganov\inst{\ref{aff20}}
\and I.~Tutusaus\orcid{0000-0002-3199-0399}\inst{\ref{aff113},\ref{aff114},\ref{aff102}}
\and J.~Valiviita\orcid{0000-0001-6225-3693}\inst{\ref{aff73},\ref{aff74}}
\and T.~Vassallo\orcid{0000-0001-6512-6358}\inst{\ref{aff2},\ref{aff57}}
\and D.~Vibert\orcid{0009-0008-0607-631X}\inst{\ref{aff39}}
\and Y.~Wang\orcid{0000-0002-4749-2984}\inst{\ref{aff62}}
\and J.~Weller\orcid{0000-0002-8282-2010}\inst{\ref{aff28},\ref{aff16}}
\and G.~Zamorani\orcid{0000-0002-2318-301X}\inst{\ref{aff22}}
\and F.~M.~Zerbi\orcid{0000-0002-9996-973X}\inst{\ref{aff21}}
\and I.~A.~Zinchenko\orcid{0000-0002-2944-2449}\inst{\ref{aff115}}
\and E.~Zucca\orcid{0000-0002-5845-8132}\inst{\ref{aff22}}
\and V.~Allevato\orcid{0000-0001-7232-5152}\inst{\ref{aff7}}
\and M.~Ballardini\orcid{0000-0003-4481-3559}\inst{\ref{aff8},\ref{aff116},\ref{aff22}}
\and P.~Bergamini\orcid{0000-0003-1383-9414}\inst{\ref{aff22}}
\and S.~Borgani\orcid{0000-0001-6151-6439}\inst{\ref{aff14},\ref{aff3},\ref{aff2},\ref{aff15},\ref{aff117}}
\and D.~R.~Boxhoorn\inst{\ref{aff9}}
\and E.~Bozzo\orcid{0000-0002-8201-1525}\inst{\ref{aff50}}
\and C.~Burigana\orcid{0000-0002-3005-5796}\inst{\ref{aff118},\ref{aff55}}
\and R.~Cabanac\orcid{0000-0001-6679-2600}\inst{\ref{aff102}}
\and M.~Calabrese\orcid{0000-0002-2637-2422}\inst{\ref{aff119},\ref{aff35}}
\and A.~Cappi\inst{\ref{aff18},\ref{aff22}}
\and F.~Caro\orcid{0009-0003-1053-0507}\inst{\ref{aff38}}
\and T.~Castro\orcid{0000-0002-6292-3228}\inst{\ref{aff120},\ref{aff2},\ref{aff3}}
\and G.~Daste\inst{\ref{aff39}}
\and B.~Dr\"oge\orcid{0000-0002-8279-868X}\inst{\ref{aff20}}
\and J.~A.~Escartin~Vigo\inst{\ref{aff16}}
\and J.~Garc\'ia-Bellido\orcid{0000-0002-9370-8360}\inst{\ref{aff101}}
\and T.~Gasparetto\orcid{0000-0002-7913-4866}\inst{\ref{aff38}}
\and V.~Gautard\inst{\ref{aff90}}
\and C.~M.~Gutierrez\orcid{0000-0001-7854-783X}\inst{\ref{aff42},\ref{aff99}}
\and J.~Macias-Perez\orcid{0000-0002-5385-2763}\inst{\ref{aff121}}
\and C.~Mancini\orcid{0000-0002-4297-0561}\inst{\ref{aff35}}
\and R.~Maoli\orcid{0000-0002-6065-3025}\inst{\ref{aff122},\ref{aff38}}
\and R.~B.~Metcalf\orcid{0000-0003-3167-2574}\inst{\ref{aff82},\ref{aff22}}
\and M.~P\"ontinen\orcid{0000-0001-5442-2530}\inst{\ref{aff73}}
\and E.~Sarpa\orcid{0000-0002-1256-655X}\inst{\ref{aff2}}
\and V.~Scottez\orcid{0009-0008-3864-940X}\inst{\ref{aff123},\ref{aff124}}
\and M.~Sereno\orcid{0000-0003-0302-0325}\inst{\ref{aff22},\ref{aff26}}
\and M.~Tenti\orcid{0000-0002-4254-5901}\inst{\ref{aff26}}
\and M.~Tucci\inst{\ref{aff50}}
\and M.~Viel\orcid{0000-0002-2642-5707}\inst{\ref{aff3},\ref{aff2},\ref{aff24},\ref{aff15},\ref{aff117}}
\and M.~Wiesmann\orcid{0009-0000-8199-5860}\inst{\ref{aff60}}
\and J.~A.~Acevedo~Barroso\orcid{0000-0002-9654-1711}\inst{\ref{aff64}}
\and Y.~Akrami\orcid{0000-0002-2407-7956}\inst{\ref{aff101},\ref{aff125}}
\and I.~T.~Andika\orcid{0000-0001-6102-9526}\inst{\ref{aff28}}
\and S.~Anselmi\orcid{0000-0002-3579-9583}\inst{\ref{aff52},\ref{aff100},\ref{aff126}}
\and M.~Archidiacono\orcid{0000-0003-4952-9012}\inst{\ref{aff78},\ref{aff79}}
\and G.~Aric\`o\orcid{0000-0002-2802-2928}\inst{\ref{aff26}}
\and F.~Atrio-Barandela\orcid{0000-0002-2130-2513}\inst{\ref{aff127}}
\and E.~Aubourg\orcid{0000-0002-5592-023X}\inst{\ref{aff86},\ref{aff128}}
\and M.~Baes\orcid{0000-0002-3930-2757}\inst{\ref{aff129}}
\and L.~Baumont\orcid{0000-0002-1518-0150}\inst{\ref{aff14},\ref{aff2},\ref{aff3}}
\and L.~Bazzanini\orcid{0000-0003-0727-0137}\inst{\ref{aff8},\ref{aff22}}
\and D.~Bertacca\orcid{0000-0002-2490-7139}\inst{\ref{aff100},\ref{aff27},\ref{aff52}}
\and M.~Bethermin\orcid{0000-0002-3915-2015}\inst{\ref{aff130}}
\and F.~Beutler\orcid{0000-0003-0467-5438}\inst{\ref{aff4}}
\and A.~Blanchard\orcid{0000-0001-8555-9003}\inst{\ref{aff102}}
\and L.~Blot\orcid{0000-0002-9622-7167}\inst{\ref{aff131},\ref{aff76}}
\and M.~L.~Brown\orcid{0000-0002-0370-8077}\inst{\ref{aff43}}
\and S.~Bruton\orcid{0000-0002-6503-5218}\inst{\ref{aff61}}
\and A.~Calabro\orcid{0000-0003-2536-1614}\inst{\ref{aff38}}
\and B.~Camacho~Quevedo\orcid{0000-0002-8789-4232}\inst{\ref{aff3},\ref{aff24},\ref{aff2}}
\and C.~S.~Carvalho\inst{\ref{aff109}}
\and F.~Cogato\orcid{0000-0003-4632-6113}\inst{\ref{aff82},\ref{aff22}}
\and T.~E.~Collett\orcid{0000-0001-5564-3140}\inst{\ref{aff132}}
\and S.~Conseil\orcid{0000-0002-3657-4191}\inst{\ref{aff133},\ref{aff45}}
\and A.~R.~Cooray\orcid{0000-0002-3892-0190}\inst{\ref{aff134}}
\and P.~Corcho-Caballero\orcid{0000-0001-6327-7080}\inst{\ref{aff9}}
\and B.~Csizi\orcid{0000-0003-3227-6581}\inst{\ref{aff105}}
\and O.~Cucciati\orcid{0000-0002-9336-7551}\inst{\ref{aff22}}
\and H.~Dannerbauer\orcid{0000-0001-7147-3575}\inst{\ref{aff42},\ref{aff99}}
\and J.~E.~Davies\orcid{0000-0002-5079-9098}\inst{\ref{aff69}}
\and S.~Davini\orcid{0000-0003-3269-1718}\inst{\ref{aff31}}
\and T.~de~Boer\orcid{0000-0001-5486-2747}\inst{\ref{aff40}}
\and G.~Desprez\orcid{0000-0001-8325-1742}\inst{\ref{aff9}}
\and A.~D\'iaz-S\'anchez\orcid{0000-0003-0748-4768}\inst{\ref{aff135}}
\and S.~Di~Domizio\orcid{0000-0003-2863-5895}\inst{\ref{aff30},\ref{aff31}}
\and J.~M.~Diego\orcid{0000-0001-9065-3926}\inst{\ref{aff136}}
\and P.-A.~Duc\orcid{0000-0003-3343-6284}\inst{\ref{aff130}}
\and F.~Dufresne\inst{\ref{aff39}}
\and V.~Duret\orcid{0009-0009-0383-4960}\inst{\ref{aff53}}
\and M.~Y.~Elkhashab\orcid{0000-0001-9306-2603}\inst{\ref{aff2},\ref{aff15},\ref{aff14},\ref{aff3}}
\and A.~Enia\orcid{0000-0002-0200-2857}\inst{\ref{aff90},\ref{aff22}}
\and Y.~Fang\orcid{0000-0002-0334-6950}\inst{\ref{aff28}}
\and A.~Farina\orcid{0009-0000-3420-929X}\inst{\ref{aff21},\ref{aff31}}
\and A.~M.~N.~Ferguson\orcid{0000-0001-7934-1278}\inst{\ref{aff4}}
\and F.~Fontanot\orcid{0000-0003-4744-0188}\inst{\ref{aff2},\ref{aff3}}
\and A.~Franco\orcid{0000-0002-4761-366X}\inst{\ref{aff137},\ref{aff138},\ref{aff139}}
\and Y.~Fu\orcid{0000-0002-0759-0504}\inst{\ref{aff63},\ref{aff9}}
\and K.~Ganga\orcid{0000-0001-8159-8208}\inst{\ref{aff86}}
\and R.~Gavazzi\orcid{0000-0002-5540-6935}\inst{\ref{aff39},\ref{aff84}}
\and E.~Gaztanaga\orcid{0000-0001-9632-0815}\inst{\ref{aff113},\ref{aff114},\ref{aff132}}
\and Z.~Ghaffari\orcid{0000-0002-6467-8078}\inst{\ref{aff2},\ref{aff3}}
\and F.~Giacomini\orcid{0000-0002-3129-2814}\inst{\ref{aff26}}
\and F.~Gianotti\orcid{0000-0003-4666-119X}\inst{\ref{aff22}}
\and G.~Gozaliasl\orcid{0000-0002-0236-919X}\inst{\ref{aff140},\ref{aff73}}
\and A.~Gruppuso\orcid{0000-0001-9272-5292}\inst{\ref{aff22},\ref{aff26}}
\and M.~Guidi\orcid{0000-0001-9408-1101}\inst{\ref{aff25},\ref{aff22}}
\and A.~Hall\orcid{0000-0002-3139-8651}\inst{\ref{aff4}}
\and N.~A.~Hatch\orcid{0000-0001-5600-0534}\inst{\ref{aff141}}
\and C.~Hern\'andez-Monteagudo\orcid{0000-0001-5471-9166}\inst{\ref{aff99},\ref{aff42}}
\and H.~Hildebrandt\orcid{0000-0002-9814-3338}\inst{\ref{aff142}}
\and J.~Hjorth\orcid{0000-0002-4571-2306}\inst{\ref{aff143}}
\and M.~Jauzac\orcid{0000-0003-1974-8732}\inst{\ref{aff144},\ref{aff83},\ref{aff145},\ref{aff146},\ref{aff102}}
\and J.~J.~E.~Kajava\orcid{0000-0002-3010-8333}\inst{\ref{aff147},\ref{aff148},\ref{aff149}}
\and Y.~Kang\orcid{0009-0000-8588-7250}\inst{\ref{aff105}}
\and V.~Kansal\orcid{0000-0002-4008-6078}\inst{\ref{aff150},\ref{aff151}}
\and D.~Karagiannis\orcid{0000-0002-4927-0816}\inst{\ref{aff8},\ref{aff152}}
\and J.~Kim\orcid{0000-0003-2776-2761}\inst{\ref{aff56}}
\and C.~C.~Kirkpatrick\inst{\ref{aff106}}
\and A.~Kov\'acs\orcid{0000-0002-5825-579X}\inst{\ref{aff153},\ref{aff154}}
\and I.~Kova{\v{c}}i{\'{c}}\orcid{0000-0001-6751-3263}\inst{\ref{aff129}}
\and K.~Koyama\orcid{0000-0001-6727-6915}\inst{\ref{aff132}}
\and M.~C.~Lam\orcid{0000-0002-9347-2298}\inst{\ref{aff4}}
\and M.~Lattanzi\orcid{0000-0003-1059-2532}\inst{\ref{aff116}}
\and M.~Lembo\orcid{0000-0002-5271-5070}\inst{\ref{aff84}}
\and G.~Leroy\orcid{0009-0004-2523-4425}\inst{\ref{aff144},\ref{aff83}}
\and G.~F.~Lesci\orcid{0000-0002-4607-2830}\inst{\ref{aff82},\ref{aff22}}
\and J.~Lesgourgues\orcid{0000-0001-7627-353X}\inst{\ref{aff155}}
\and T.~I.~Liaudat\orcid{0000-0002-9104-314X}\inst{\ref{aff128}}
\and L.~Linke\orcid{0000-0002-2622-8113}\inst{\ref{aff105}}
\and S.~J.~Liu\orcid{0000-0001-7680-2139}\inst{\ref{aff54}}
\and G.~Maggio\orcid{0000-0003-4020-4836}\inst{\ref{aff2}}
\and F.~Mannucci\orcid{0000-0002-4803-2381}\inst{\ref{aff156}}
\and F.~R.~Marleau\orcid{0000-0002-1442-2947}\inst{\ref{aff105}}
\and C.~J.~A.~P.~Martins\orcid{0000-0002-4886-9261}\inst{\ref{aff157},\ref{aff158}}
\and M.~Migliaccio\inst{\ref{aff159},\ref{aff160}}
\and M.~Miluzio\inst{\ref{aff6},\ref{aff161}}
\and G.~Morgante\inst{\ref{aff22}}
\and S.~Nadathur\orcid{0000-0001-9070-3102}\inst{\ref{aff132}}
\and K.~Naidoo\orcid{0000-0002-9182-1802}\inst{\ref{aff132},\ref{aff69}}
\and A.~Navarro-Alsina\orcid{0000-0002-3173-2592}\inst{\ref{aff80}}
\and S.~Nesseris\orcid{0000-0002-0567-0324}\inst{\ref{aff101}}
\and L.~Nicastro\orcid{0000-0001-8534-6788}\inst{\ref{aff22}}
\and F.~Oppizzi\orcid{0000-0003-3904-8370}\inst{\ref{aff31},\ref{aff52},\ref{aff100}}
\and F.~Pace\orcid{0000-0001-8039-0480}\inst{\ref{aff32},\ref{aff33},\ref{aff34},\ref{aff162}}
\and D.~Paoletti\orcid{0000-0003-4761-6147}\inst{\ref{aff22},\ref{aff55}}
\and F.~Passalacqua\orcid{0000-0002-8606-4093}\inst{\ref{aff52}}
\and K.~Paterson\orcid{0000-0001-8340-3486}\inst{\ref{aff69}}
\and C.~Pattison\orcid{0000-0003-3272-2617}\inst{\ref{aff132}}
\and R.~Pello\orcid{0000-0003-0858-6109}\inst{\ref{aff39}}
\and A.~Pisani\orcid{0000-0002-6146-4437}\inst{\ref{aff53}}
\and D.~Potter\orcid{0000-0002-0757-5195}\inst{\ref{aff163}}
\and G.~W.~Pratt\inst{\ref{aff5}}
\and S.~Quai\orcid{0000-0002-0449-8163}\inst{\ref{aff82},\ref{aff22}}
\and M.~Radovich\orcid{0000-0002-3585-866X}\inst{\ref{aff27}}
\and G.~Rodighiero\orcid{0000-0002-9415-2296}\inst{\ref{aff100},\ref{aff27}}
\and W.~Roster\orcid{0000-0002-9149-6528}\inst{\ref{aff16}}
\and M.~Sahl\'en\orcid{0000-0003-0973-4804}\inst{\ref{aff164}}
\and D.~B.~Sanders\orcid{0000-0002-1233-9998}\inst{\ref{aff40}}
\and A.~Schneider\orcid{0000-0001-7055-8104}\inst{\ref{aff163}}
\and D.~Sciotti\orcid{0009-0008-4519-2620}\inst{\ref{aff38},\ref{aff81}}
\and E.~Sellentin\orcid{0009-0002-2655-3458}\inst{\ref{aff165},\ref{aff63}}
\and S.~Serjeant\orcid{0000-0002-0517-7943}\inst{\ref{aff166}}
\and F.~Shankar\orcid{0000-0001-8973-5051}\inst{\ref{aff167}}
\and A.~Shulevski\orcid{0000-0002-1827-0469}\inst{\ref{aff168},\ref{aff9},\ref{aff169}}
\and L.~C.~Smith\orcid{0000-0002-3259-2771}\inst{\ref{aff170}}
\and F.~Soldano\inst{\ref{aff84}}
\and E.~Soubrie\orcid{0000-0001-9295-1863}\inst{\ref{aff51}}
\and I.~Szapudi\orcid{0000-0003-2274-0301}\inst{\ref{aff40}}
\and M.~Talia\orcid{0000-0003-4352-2063}\inst{\ref{aff82},\ref{aff22}}
\and K.~Tanidis\orcid{0000-0001-9843-5130}\inst{\ref{aff171}}
\and C.~Tao\orcid{0000-0001-7961-8177}\inst{\ref{aff53}}
\and F.~Tarsitano\orcid{0000-0002-5919-0238}\inst{\ref{aff172},\ref{aff173},\ref{aff50}}
\and G.~Testera\orcid{0000-0003-2970-766X}\inst{\ref{aff31}}
\and R.~Teyssier\orcid{0000-0001-7689-0933}\inst{\ref{aff174}}
\and S.~Tosi\orcid{0000-0002-7275-9193}\inst{\ref{aff30},\ref{aff21},\ref{aff31}}
\and A.~Troja\orcid{0000-0003-0239-4595}\inst{\ref{aff2}}
\and C.~Uhlemann\orcid{0000-0001-7831-1579}\inst{\ref{aff175},\ref{aff176}}
\and M.~Urbano\orcid{0000-0001-5640-0650}\inst{\ref{aff141},\ref{aff130}}
\and C.~Valieri\inst{\ref{aff26}}
\and A.~Venhola\orcid{0000-0001-6071-4564}\inst{\ref{aff177}}
\and D.~Vergani\orcid{0000-0003-0898-2216}\inst{\ref{aff22}}
\and E.~Vilenius\orcid{0000-0002-6184-7681}\inst{\ref{aff23}}
\and S.~Vinciguerra\orcid{0009-0005-4018-3184}\inst{\ref{aff39}}
\and M.~von~Wietersheim-Kramsta\orcid{0000-0003-4986-5091}\inst{\ref{aff83},\ref{aff144}}
\and N.~A.~Walton\orcid{0000-0003-3983-8778}\inst{\ref{aff170}}
\and L.~Wang\orcid{0000-0002-6736-9158}\inst{\ref{aff178},\ref{aff9}}
\and A.~H.~Wright\orcid{0000-0001-7363-7932}\inst{\ref{aff142}}
\and H.~Boquet\inst{\ref{aff75}}
\and D.~Scott\orcid{0000-0002-6878-9840}\inst{\ref{aff179}}}
										   
\institute{Centre National d'Etudes Spatiales -- Centre spatial de Toulouse, 18 avenue Edouard Belin, 31401 Toulouse Cedex 9, France\label{aff1}
\and
INAF-Osservatorio Astronomico di Trieste, Via G. B. Tiepolo 11, 34143 Trieste, Italy\label{aff2}
\and
IFPU, Institute for Fundamental Physics of the Universe, via Beirut 2, 34151 Trieste, Italy\label{aff3}
\and
Institute for Astronomy, University of Edinburgh, Royal Observatory, Blackford Hill, Edinburgh EH9 3HJ, UK\label{aff4}
\and
Universit\'e Paris-Saclay, Universit\'e Paris Cit\'e, CEA, CNRS, AIM, 91191, Gif-sur-Yvette, France\label{aff5}
\and
ESAC/ESA, Camino Bajo del Castillo, s/n., Urb. Villafranca del Castillo, 28692 Villanueva de la Ca\~nada, Madrid, Spain\label{aff6}
\and
INAF-Osservatorio Astronomico di Capodimonte, Via Moiariello 16, 80131 Napoli, Italy\label{aff7}
\and
Dipartimento di Fisica e Scienze della Terra, Universit\`a degli Studi di Ferrara, Via Giuseppe Saragat 1, 44122 Ferrara, Italy\label{aff8}
\and
Kapteyn Astronomical Institute, University of Groningen, PO Box 800, 9700 AV Groningen, The Netherlands\label{aff9}
\and
ATG Europe BV, Huygensstraat 34, 2201 DK Noordwijk, The Netherlands\label{aff10}
\and
Thales~Services~S.A.S., 290 All\'ee du Lac, 31670 Lab\`ege, France\label{aff11}
\and
Department of Physics "E. Pancini", University Federico II, Via Cinthia 6, 80126, Napoli, Italy\label{aff12}
\and
INFN -- Sezione di Napoli, Via Cinthia 6, 80126, Napoli, Italy\label{aff13}
\and
Dipartimento di Fisica - Sezione di Astronomia, Universit\`a di Trieste, Via Tiepolo 11, 34131 Trieste, Italy\label{aff14}
\and
INFN, Sezione di Trieste, Via Valerio 2, 34127 Trieste TS, Italy\label{aff15}
\and
Max Planck Institute for Extraterrestrial Physics, Giessenbachstr. 1, 85748 Garching, Germany\label{aff16}
\and
University of Applied Sciences and Arts of Northwestern Switzerland, School of Computer Science, 5210 Windisch, Switzerland\label{aff17}
\and
Universit\'e C\^{o}te d'Azur, Observatoire de la C\^{o}te d'Azur, CNRS, Laboratoire Lagrange, Bd de l'Observatoire, CS 34229, 06304 Nice cedex 4, France\label{aff18}
\and
University of Applied Sciences and Arts of Northwestern Switzerland, School of Engineering, 5210 Windisch, Switzerland\label{aff19}
\and
Centre for Information Technology, University of Groningen, P.O. Box 11044, 9700 CA Groningen, The Netherlands\label{aff20}
\and
INAF-Osservatorio Astronomico di Brera, Via Brera 28, 20122 Milano, Italy\label{aff21}
\and
INAF-Osservatorio di Astrofisica e Scienza dello Spazio di Bologna, Via Piero Gobetti 93/3, 40129 Bologna, Italy\label{aff22}
\and
Mullard Space Science Laboratory, University College London, Holmbury St Mary, Dorking, Surrey RH5 6NT, UK\label{aff23}
\and
SISSA, International School for Advanced Studies, Via Bonomea 265, 34136 Trieste TS, Italy\label{aff24}
\and
Dipartimento di Fisica e Astronomia, Universit\`a di Bologna, Via Gobetti 93/2, 40129 Bologna, Italy\label{aff25}
\and
INFN-Sezione di Bologna, Viale Berti Pichat 6/2, 40127 Bologna, Italy\label{aff26}
\and
INAF-Osservatorio Astronomico di Padova, Via dell'Osservatorio 5, 35122 Padova, Italy\label{aff27}
\and
Universit\"ats-Sternwarte M\"unchen, Fakult\"at f\"ur Physik, Ludwig-Maximilians-Universit\"at M\"unchen, Scheinerstr.~1, 81679 M\"unchen, Germany\label{aff28}
\and
Space Science Data Center, Italian Space Agency, via del Politecnico snc, 00133 Roma, Italy\label{aff29}
\and
Dipartimento di Fisica, Universit\`a di Genova, Via Dodecaneso 33, 16146, Genova, Italy\label{aff30}
\and
INFN-Sezione di Genova, Via Dodecaneso 33, 16146, Genova, Italy\label{aff31}
\and
Dipartimento di Fisica, Universit\`a degli Studi di Torino, Via P. Giuria 1, 10125 Torino, Italy\label{aff32}
\and
INFN-Sezione di Torino, Via P. Giuria 1, 10125 Torino, Italy\label{aff33}
\and
INAF-Osservatorio Astrofisico di Torino, Via Osservatorio 20, 10025 Pino Torinese (TO), Italy\label{aff34}
\and
INAF-IASF Milano, Via Alfonso Corti 12, 20133 Milano, Italy\label{aff35}
\and
Centro de Investigaciones Energ\'eticas, Medioambientales y Tecnol\'ogicas (CIEMAT), Avenida Complutense 40, 28040 Madrid, Spain\label{aff36}
\and
Port d'Informaci\'{o} Cient\'{i}fica, Campus UAB, C. Albareda s/n, 08193 Bellaterra (Barcelona), Spain\label{aff37}
\and
INAF-Osservatorio Astronomico di Roma, Via Frascati 33, 00078 Monteporzio Catone, Italy\label{aff38}
\and
Aix-Marseille Universit\'e, CNRS, CNES, LAM, Marseille, France\label{aff39}
\and
Institute for Astronomy, University of Hawaii, 2680 Woodlawn Drive, Honolulu, HI 96822, USA\label{aff40}
\and
Dipartimento di Fisica e Astronomia "Augusto Righi" - Alma Mater Studiorum Universit\`a di Bologna, Viale Berti Pichat 6/2, 40127 Bologna, Italy\label{aff41}
\and
Instituto de Astrof\'{\i}sica de Canarias, E-38205 La Laguna, Tenerife, Spain\label{aff42}
\and
Jodrell Bank Centre for Astrophysics, Department of Physics and Astronomy, University of Manchester, Oxford Road, Manchester M13 9PL, UK\label{aff43}
\and
European Space Agency/ESRIN, Largo Galileo Galilei 1, 00044 Frascati, Roma, Italy\label{aff44}
\and
Universit\'e Lyon 1, CNRS/IN2P3, IP2I Lyon, Villeurbanne, France\label{aff45}
\and
Institut de Ci\`{e}ncies del Cosmos (ICCUB), Universitat de Barcelona (IEEC-UB), Mart\'{i} i Franqu\`{e}s 1, 08028 Barcelona, Spain\label{aff46}
\and
Instituci\'o Catalana de Recerca i Estudis Avan\c{c}ats (ICREA), Passeig de Llu\'{\i}s Companys 23, 08010 Barcelona, Spain\label{aff47}
\and
Institut de Ciencies de l'Espai (IEEC-CSIC), Campus UAB, Carrer de Can Magrans, s/n Cerdanyola del Vall\'es, 08193 Barcelona, Spain\label{aff48}
\and
Universit\'e Lyon 1, CNRS/IN2P3, IUF, IP2I Lyon, Villeurbanne, France \label{aff49}
\and
Department of Astronomy, University of Geneva, ch. d'Ecogia 16, 1290 Versoix, Switzerland\label{aff50}
\and
Universit\'e Paris-Saclay, CNRS, Institut d'astrophysique spatiale, 91405, Orsay, France\label{aff51}
\and
INFN-Padova, Via Marzolo 8, 35131 Padova, Italy\label{aff52}
\and
Aix-Marseille Universit\'e, CNRS/IN2P3, CPPM, Marseille, France\label{aff53}
\and
INAF-Istituto di Astrofisica e Planetologia Spaziali, via del Fosso del Cavaliere, 100, 00100 Roma, Italy\label{aff54}
\and
INFN-Bologna, Via Irnerio 46, 40126 Bologna, Italy\label{aff55}
\and
Department of Physics, University of Oxford, Keble Road, Oxford OX1 3RH, UK\label{aff56}
\and
University Observatory, LMU Faculty of Physics, Scheinerstr.~1, 81679 Munich, Germany\label{aff57}
\and
FRACTAL S.L.N.E., calle Tulip\'an 2, Portal 13 1A, 28231, Las Rozas de Madrid, Spain\label{aff58}
\and
National Research Council, Herzberg Astronomy and Astrophysics Research Centre, 5071 W. Saanich Rd. Victoria, BC, V9E 2E7, Canada\label{aff59}
\and
Institute of Theoretical Astrophysics, University of Oslo, P.O. Box 1029 Blindern, 0315 Oslo, Norway\label{aff60}
\and
California Institute of Technology, 1200 E California Blvd, Pasadena, CA 91125, USA\label{aff61}
\and
Caltech/IPAC, 1200 E. California Blvd., Pasadena, CA 91125, USA\label{aff62}
\and
Leiden Observatory, Leiden University, Einsteinweg 55, 2333 CC Leiden, The Netherlands\label{aff63}
\and
Jet Propulsion Laboratory, California Institute of Technology, 4800 Oak Grove Drive, Pasadena, CA, 91109, USA\label{aff64}
\and
Department of Physics, Lancaster University, Lancaster, LA1 4YB, UK\label{aff65}
\and
Felix Hormuth Engineering, Goethestr. 17, 69181 Leimen, Germany\label{aff66}
\and
Technical University of Denmark, Elektrovej 327, 2800 Kgs. Lyngby, Denmark\label{aff67}
\and
Cosmic Dawn Center (DAWN), Denmark\label{aff68}
\and
Max-Planck-Institut f\"ur Astronomie, K\"onigstuhl 17, 69117 Heidelberg, Germany\label{aff69}
\and
NASA Goddard Space Flight Center, Greenbelt, MD 20771, USA\label{aff70}
\and
Department of Physics and Astronomy, University College London, Gower Street, London WC1E 6BT, UK\label{aff71}
\and
Universit\'e de Gen\`eve, D\'epartement de Physique Th\'eorique and Centre for Astroparticle Physics, 24 quai Ernest-Ansermet, CH-1211 Gen\`eve 4, Switzerland\label{aff72}
\and
Department of Physics, P.O. Box 64, University of Helsinki, 00014 Helsinki, Finland\label{aff73}
\and
Helsinki Institute of Physics, Gustaf H{\"a}llstr{\"o}min katu 2, University of Helsinki, 00014 Helsinki, Finland\label{aff74}
\and
Centre de Calcul de l'IN2P3/CNRS, 21 avenue Pierre de Coubertin 69627 Villeurbanne Cedex, France\label{aff75}
\and
Laboratoire d'etude de l'Univers et des phenomenes eXtremes, Observatoire de Paris, Universit\'e PSL, Sorbonne Universit\'e, CNRS, 92190 Meudon, France\label{aff76}
\and
SKAO, Jodrell Bank, Lower Withington, Macclesfield SK11 9FT, UK\label{aff77}
\and
Dipartimento di Fisica "Aldo Pontremoli", Universit\`a degli Studi di Milano, Via Celoria 16, 20133 Milano, Italy\label{aff78}
\and
INFN-Sezione di Milano, Via Celoria 16, 20133 Milano, Italy\label{aff79}
\and
Universit\"at Bonn, Argelander-Institut f\"ur Astronomie, Auf dem H\"ugel 71, 53121 Bonn, Germany\label{aff80}
\and
INFN-Sezione di Roma, Piazzale Aldo Moro, 2 - c/o Dipartimento di Fisica, Edificio G. Marconi, 00185 Roma, Italy\label{aff81}
\and
Dipartimento di Fisica e Astronomia "Augusto Righi" - Alma Mater Studiorum Universit\`a di Bologna, via Piero Gobetti 93/2, 40129 Bologna, Italy\label{aff82}
\and
Department of Physics, Institute for Computational Cosmology, Durham University, South Road, Durham, DH1 3LE, UK\label{aff83}
\and
Institut d'Astrophysique de Paris, UMR 7095, CNRS, and Sorbonne Universit\'e, 98 bis boulevard Arago, 75014 Paris, France\label{aff84}
\and
Niels Bohr Institute, University of Copenhagen, Jagtvej 128, 2200 Copenhagen, Denmark\label{aff85}
\and
Universit\'e Paris Cit\'e, CNRS, Astroparticule et Cosmologie, 75013 Paris, France\label{aff86}
\and
CNRS-UCB International Research Laboratory, Centre Pierre Bin\'etruy, IRL2007, CPB-IN2P3, Berkeley, USA\label{aff87}
\and
Institute of Physics, Laboratory of Astrophysics, Ecole Polytechnique F\'ed\'erale de Lausanne (EPFL), Observatoire de Sauverny, 1290 Versoix, Switzerland\label{aff88}
\and
Telespazio UK S.L. for European Space Agency (ESA), Camino bajo del Castillo, s/n, Urbanizacion Villafranca del Castillo, Villanueva de la Ca\~nada, 28692 Madrid, Spain\label{aff89}
\and
CEA Saclay, DFR/IRFU, Service d'Astrophysique, Bat. 709, 91191 Gif-sur-Yvette, France\label{aff90}
\and
Institut de F\'{i}sica d'Altes Energies (IFAE), The Barcelona Institute of Science and Technology, Campus UAB, 08193 Bellaterra (Barcelona), Spain\label{aff91}
\and
School of Mathematics and Physics, University of Surrey, Guildford, Surrey, GU2 7XH, UK\label{aff92}
\and
European Space Agency/ESTEC, Keplerlaan 1, 2201 AZ Noordwijk, The Netherlands\label{aff93}
\and
Waterloo Centre for Astrophysics, University of Waterloo, Waterloo, Ontario N2L 3G1, Canada\label{aff94}
\and
Department of Physics and Astronomy, University of Waterloo, Waterloo, Ontario N2L 3G1, Canada\label{aff95}
\and
Perimeter Institute for Theoretical Physics, Waterloo, Ontario N2L 2Y5, Canada\label{aff96}
\and
Institute of Space Science, Str. Atomistilor, nr. 409 M\u{a}gurele, Ilfov, 077125, Romania\label{aff97}
\and
Consejo Superior de Investigaciones Cientificas, Calle Serrano 117, 28006 Madrid, Spain\label{aff98}
\and
Universidad de La Laguna, Dpto. Astrof\'\i sica, E-38206 La Laguna, Tenerife, Spain\label{aff99}
\and
Dipartimento di Fisica e Astronomia "G. Galilei", Universit\`a di Padova, Via Marzolo 8, 35131 Padova, Italy\label{aff100}
\and
Instituto de F\'isica Te\'orica UAM-CSIC, Campus de Cantoblanco, 28049 Madrid, Spain\label{aff101}
\and
Institut de Recherche en Astrophysique et Plan\'etologie (IRAP), Universit\'e de Toulouse, CNRS, UPS, CNES, 14 Av. Edouard Belin, 31400 Toulouse, France\label{aff102}
\and
Universit\'e St Joseph; Faculty of Sciences, Beirut, Lebanon\label{aff103}
\and
Departamento de F\'isica, FCFM, Universidad de Chile, Blanco Encalada 2008, Santiago, Chile\label{aff104}
\and
Universit\"at Innsbruck, Institut f\"ur Astro- und Teilchenphysik, Technikerstr. 25/8, 6020 Innsbruck, Austria\label{aff105}
\and
Department of Physics and Helsinki Institute of Physics, Gustaf H\"allstr\"omin katu 2, University of Helsinki, 00014 Helsinki, Finland\label{aff106}
\and
Infrared Processing and Analysis Center, California Institute of Technology, Pasadena, CA 91125, USA\label{aff107}
\and
Departamento de F\'isica, Faculdade de Ci\^encias, Universidade de Lisboa, Edif\'icio C8, Campo Grande, PT1749-016 Lisboa, Portugal\label{aff108}
\and
Instituto de Astrof\'isica e Ci\^encias do Espa\c{c}o, Faculdade de Ci\^encias, Universidade de Lisboa, Tapada da Ajuda, 1349-018 Lisboa, Portugal\label{aff109}
\and
Cosmic Dawn Center (DAWN)\label{aff110}
\and
Universidad Polit\'ecnica de Cartagena, Departamento de Electr\'onica y Tecnolog\'ia de Computadoras,  Plaza del Hospital 1, 30202 Cartagena, Spain\label{aff111}
\and
European University of Technology EUt+, European Union\label{aff112}
\and
Institute of Space Sciences (ICE, CSIC), Campus UAB, Carrer de Can Magrans, s/n, 08193 Barcelona, Spain\label{aff113}
\and
Institut d'Estudis Espacials de Catalunya (IEEC),  Edifici RDIT, Campus UPC, 08860 Castelldefels, Barcelona, Spain\label{aff114}
\and
Astronomisches Rechen-Institut, Zentrum f\"ur Astronomie der Universit\"at Heidelberg, M\"onchhofstr. 12-14, 69120 Heidelberg, Germany\label{aff115}
\and
Istituto Nazionale di Fisica Nucleare, Sezione di Ferrara, Via Giuseppe Saragat 1, 44122 Ferrara, Italy\label{aff116}
\and
ICSC - Centro Nazionale di Ricerca in High Performance Computing, Big Data e Quantum Computing, Via Magnanelli 2, Bologna, Italy\label{aff117}
\and
INAF, Istituto di Radioastronomia, Via Piero Gobetti 101, 40129 Bologna, Italy\label{aff118}
\and
Astronomical Observatory of the Autonomous Region of the Aosta Valley (OAVdA), Loc. Lignan 39, I-11020, Nus (Aosta Valley), Italy\label{aff119}
\and
Department of Mathematical Physics, Institute of Physics, University of S\~ao Paulo, R. do Mat\~ao 1371, 05508-090, S\~ao Paulo, SP, Brazil\label{aff120}
\and
Univ. Grenoble Alpes, CNRS, Grenoble INP, LPSC-IN2P3, 53, Avenue des Martyrs, 38000 Grenoble, France\label{aff121}
\and
Dipartimento di Fisica, Sapienza Universit\`a di Roma, Piazzale Aldo Moro 2, 00185 Roma, Italy\label{aff122}
\and
Institut d'Astrophysique de Paris, 98bis Boulevard Arago, 75014, Paris, France\label{aff123}
\and
ICL, Junia, Universit\'e Catholique de Lille, LITL, 59000 Lille, France\label{aff124}
\and
CERCA/ISO, Department of Physics, Case Western Reserve University, 10900 Euclid Avenue, Cleveland, OH 44106, USA\label{aff125}
\and
Laboratoire Univers et Th\'eorie, Observatoire de Paris, Universit\'e PSL, Universit\'e Paris Cit\'e, CNRS, 92190 Meudon, France\label{aff126}
\and
Departamento de F{\'\i}sica Fundamental. Universidad de Salamanca. Plaza de la Merced s/n. 37008 Salamanca, Spain\label{aff127}
\and
IRFU, CEA, Universit\'e Paris-Saclay 91191 Gif-sur-Yvette Cedex, France\label{aff128}
\and
Universiteit Gent, Department of Physics and Astronomy, Proeftuinstraat 86 N3, 9000 Ghent, Belgium
\label{aff129}
\and
Universit\'e de Strasbourg, CNRS, Observatoire astronomique de Strasbourg, UMR 7550, 67000 Strasbourg, France\label{aff130}
\and
Center for Data-Driven Discovery, Kavli IPMU (WPI), UTIAS, The University of Tokyo, Kashiwa, Chiba 277-8583, Japan\label{aff131}
\and
Institute of Cosmology and Gravitation, University of Portsmouth, Portsmouth PO1 3FX, UK\label{aff132}
\and
Centre de Recherche Astrophysique de Lyon, UMR5574, CNRS, Universit\'e Claude Bernard Lyon 1, ENS de Lyon, 69230, Saint-Genis-Laval, France\label{aff133}
\and
Department of Physics \& Astronomy, University of California Irvine, Irvine CA 92697, USA\label{aff134}
\and
Departamento F\'isica Aplicada, Universidad Polit\'ecnica de Cartagena, Campus Muralla del Mar, 30202 Cartagena, Murcia, Spain\label{aff135}
\and
Instituto de F\'isica de Cantabria, Edificio Juan Jord\'a, Avenida de los Castros, 39005 Santander, Spain\label{aff136}
\and
Department of Mathematics and Physics E. De Giorgi, University of Salento, Via per Arnesano, CP-I93, 73100, Lecce, Italy\label{aff137}
\and
INFN, Sezione di Lecce, Via per Arnesano, CP-193, 73100, Lecce, Italy\label{aff138}
\and
INAF-Sezione di Lecce, c/o Dipartimento Matematica e Fisica, Via per Arnesano, 73100, Lecce, Italy\label{aff139}
\and
Department of Computer Science, Aalto University, PO Box 15400, Espoo, FI-00 076, Finland\label{aff140}
\and
School of Physics and Astronomy, University of Nottingham, University Park, Nottingham NG7 2RD, UK\label{aff141}
\and
Ruhr University Bochum, Faculty of Physics and Astronomy, Astronomical Institute (AIRUB), German Centre for Cosmological Lensing (GCCL), 44780 Bochum, Germany\label{aff142}
\and
DARK, Niels Bohr Institute, University of Copenhagen, Jagtvej 155, 2200 Copenhagen, Denmark\label{aff143}
\and
Department of Physics, Centre for Extragalactic Astronomy, Durham University, South Road, Durham, DH1 3LE, UK\label{aff144}
\and
Astrophysics Research Centre, University of KwaZulu-Natal, Westville Campus, Durban 4041, South Africa\label{aff145}
\and
School of Mathematics, Statistics \& Computer Science, University of KwaZulu-Natal, Westville Campus, Durban 4041, South Africa\label{aff146}
\and
Department of Physics and Astronomy, Vesilinnantie 5, University of Turku, 20014 Turku, Finland\label{aff147}
\and
Finnish Centre for Astronomy with ESO (FINCA), Quantum, Vesilinnantie 5, University of Turku, 20014 Turku, Finland\label{aff148}
\and
Serco for European Space Agency (ESA), Camino bajo del Castillo, s/n, Urbanizacion Villafranca del Castillo, Villanueva de la Ca\~nada, 28692 Madrid, Spain\label{aff149}
\and
ARC Centre of Excellence for Dark Matter Particle Physics, Melbourne, Australia\label{aff150}
\and
Centre for Astrophysics \& Supercomputing, Swinburne University of Technology,  Hawthorn, Victoria 3122, Australia\label{aff151}
\and
Department of Physics and Astronomy, University of the Western Cape, Bellville, Cape Town, 7535, South Africa\label{aff152}
\and
MTA-CSFK Lend\"ulet Large-Scale Structure Research Group, Konkoly-Thege Mikl\'os \'ut 15-17, H-1121 Budapest, Hungary\label{aff153}
\and
Konkoly Observatory, HUN-REN CSFK, MTA Centre of Excellence, Budapest, Konkoly Thege Mikl\'os {\'u}t 15-17. H-1121, Hungary\label{aff154}
\and
Institute for Theoretical Particle Physics and Cosmology (TTK), RWTH Aachen University, 52056 Aachen, Germany\label{aff155}
\and
INAF-Osservatorio Astrofisico di Arcetri, Largo E. Fermi 5, 50125, Firenze, Italy\label{aff156}
\and
Centro de Astrof\'{\i}sica da Universidade do Porto, Rua das Estrelas, 4150-762 Porto, Portugal\label{aff157}
\and
Instituto de Astrof\'isica e Ci\^encias do Espa\c{c}o, Universidade do Porto, CAUP, Rua das Estrelas, PT4150-762 Porto, Portugal\label{aff158}
\and
Dipartimento di Fisica, Universit\`a di Roma Tor Vergata, Via della Ricerca Scientifica 1, Roma, Italy\label{aff159}
\and
INFN, Sezione di Roma 2, Via della Ricerca Scientifica 1, Roma, Italy\label{aff160}
\and
HE Space for European Space Agency (ESA), Camino bajo del Castillo, s/n, Urbanizacion Villafranca del Castillo, Villanueva de la Ca\~nada, 28692 Madrid, Spain\label{aff161}
\and
Instituto de Astrof\'isica e Ci\^encias do Espa\c{c}o, Faculdade de Ci\^encias, Universidade de Lisboa, Campo Grande, 1749-016 Lisboa, Portugal\label{aff162}
\and
Department of Astrophysics, University of Zurich, Winterthurerstrasse 190, 8057 Zurich, Switzerland\label{aff163}
\and
Theoretical astrophysics, Department of Physics and Astronomy, Uppsala University, Box 524, 751 20 Uppsala, Sweden\label{aff164}
\and
Mathematical Institute, University of Leiden, Einsteinweg 55, 2333 CA Leiden, The Netherlands\label{aff165}
\and
School of Physical Sciences, The Open University, Milton Keynes, MK7 6AA, UK\label{aff166}
\and
School of Physics \& Astronomy, University of Southampton, Highfield Campus, Southampton SO17 1BJ, UK\label{aff167}
\and
ASTRON, the Netherlands Institute for Radio Astronomy, Postbus 2, 7990 AA, Dwingeloo, The Netherlands\label{aff168}
\and
Center for Advanced Interdisciplinary Research, Ss. Cyril and Methodius University in Skopje, Macedonia\label{aff169}
\and
Institute of Astronomy, University of Cambridge, Madingley Road, Cambridge CB3 0HA, UK\label{aff170}
\and
Center for Astrophysics and Cosmology, University of Nova Gorica, Nova Gorica, Slovenia\label{aff171}
\and
Kobayashi-Maskawa Institute for the Origin of Particles and the Universe, Nagoya University, Chikusa-ku, Nagoya, 464-8602, Japan\label{aff172}
\and
Institute for Particle Physics and Astrophysics, Dept. of Physics, ETH Zurich, Wolfgang-Pauli-Strasse 27, 8093 Zurich, Switzerland\label{aff173}
\and
Department of Astrophysical Sciences, Peyton Hall, Princeton University, Princeton, NJ 08544, USA\label{aff174}
\and
Fakult\"at f\"ur Physik, Universit\"at Bielefeld, Postfach 100131, 33501 Bielefeld, Germany\label{aff175}
\and
School of Mathematics, Statistics and Physics, Newcastle University, Herschel Building, Newcastle-upon-Tyne, NE1 7RU, UK\label{aff176}
\and
Space physics and astronomy research unit, University of Oulu, Pentti Kaiteran katu 1, FI-90014 Oulu, Finland\label{aff177}
\and
SRON Netherlands Institute for Space Research, Landleven 12, 9747 AD, Groningen, The Netherlands\label{aff178}
\and
Department of Physics and Astronomy, University of British Columbia, Vancouver, BC V6T 1Z1, Canada\label{aff179}}